\documentclass[fleqn,usenatbib]{mnras}
\usepackage[T1]{fontenc}

\DeclareRobustCommand{\VAN}[3]{#2}
\let\VANthebibliography\thebibliography
\def\thebibliography{\DeclareRobustCommand{\VAN}[3]{##3}\VANthebibliography}

\providecommand{\putpic}[2][1]{\includegraphics[width=#1\linewidth]{plots/#2}}
\providecommand{\nopic}[1]{{\color{red} Image: \detokenize{plots/#1} not found!}}
\providecommand{\pic}[2][1]{ \IfFileExists{plots/#2.pdf}{\putpic[#1]{#2}}{\IfFileExists{plots/#2.png}{ \putpic[#1]{#2} } { \IfFileExists{plots/#2.jpg}{\putpic[#1]{#2}}{\nopic{#2}} } } }

\usepackage{amsmath,amssymb,amsfonts}
\usepackage{url}
\usepackage{newtxtext,newtxmath}

\usepackage{tikz}
\usetikzlibrary{shapes, arrows, shadows}
\usetikzlibrary{decorations.pathreplacing}

\providecommand{\pic}[2][1]{\includegraphics[width=#1\linewidth]{plots/#2}}

\providecommand{\nolinkurl}[1]{\url{#1}}

\newcommand{\code}[1]{\texttt{\detokenize{#1}}}
\title[Reduction pipeline for TNO occultations]{A reduction procedure and pipeline for the detection of trans-Neptunian objects using occultations}

\author[G. Nir et al.]{
Guy Nir,$^{1,2}$\thanks{E-mail: guyn@berkeley.edu (GN)}
Eran O.~Ofek,$^{1}$
and
Barak Zackay$^{1}$
\\
$^{1}$Department of Particle Physics and astrophysics, the Weizmann Institute of Science, Rehovot, Israel\\
$^{2}$University of California, Berkeley, Department of Astronomy, Berkeley, CA 94720
}

\date{Accepted 17/08/2023. Received 14/08/2023; in original form 04/04/2023}

\pubyear{2023}

\begin{document}
\label{firstpage}
\pagerange{\pageref{firstpage}--\pageref{lastpage}}
\maketitle

	
\begin{abstract}
    Kuiper belt objects smaller than a few kilometers are difficult to observe directly.
    They can be detected when they randomly occult a background star.
    Close to the ecliptic plane, each star is occulted once every tens of thousands of hours,
    and occultations typically last for less than a second.
    We present an algorithm, and companion pipeline,
    for detection of diffractive occultation events.
    Our approach includes: cleaning the data; an efficient and optimal matched filtering of the light-curves
    with a template bank of diffractive occultations; treating the red-noise in the light-curves;
    injection of simulated events for efficiency estimation; and applying data quality cuts.
    We discuss human vetting of the candidate events in a blinded way to
    reduce bias caused by the human-in-the-loop.
    We present Markov Chain Monte Carlo tools to estimate the parameters of candidate occultations,
    and test them on simulated events.
    This pipeline is used by the Weizmann Fast Astronomical Survey Telescope (W-FAST).
    The methods discussed here can be applied to searches for other Trans-Neptunian objects, 
    albeit with larger radii that correspond to a larger diffraction length-scale. 
\end{abstract}

	
	
\section{Introduction}
    
    The Kuiper belt is a collection of small bodies 
    orbiting the Sun outside the orbit of Neptune, 
    mostly with small inclinations relative to the ecliptic plane. 
    These objects presumably formed in the early stages of 
    the Solar System's formation. 
    Close to the Sun, such planetesimals were ejected or accreted
    by the forming and migrating planets, 
    but outside the planetary region a distribution 
    of icy bodies remain, 
    holding clues about the dynamical history of the Solar System. 
    Kuiper Belt Objects (KBOs) are a possible 
    source of the Jupiter Family Comets 
    \citep{kuiper_belt_jupiter_family_Levison_1997,scattered_disk_jupiter_family_Duncan_1997,scattered_disk_jupiter_family_Volk_2008, KBO_Jupiter_family_Wesley_2022}.
    In addition, 
    studying the collisional history of the Kuiper belt
    can inform us of the mechanisms for dust production 
    in our Solar System, 
    that is also visible around other stars
    (e.g., \citealt{kuiper_belt_other_stars_eta_corvi_Wyatt_2005}).
	
    The size distribution of KBOs is well 
    described by a power-law (where $N$ is the number of objects above a radius $r$): 
    \begin{equation}
        N(>r) \propto r^{1-q}, 
    \end{equation}
    albeit with different slopes at different parts of the size distribution. 
    Above $r=40$\,km the power law index is measured to be $q\sim 4.5$,
    \citep{kuiper_belt_survey_HST_Bernstein_2004,kuiper_belt_survey_Subaru_archival_Fuentes_2008,kuiper_belt_survey_Subaru_Fraser_2009}.
    Below 40\,km the power-law index is estimated to be $q\sim 3.8$ \citep{abundance_kuiper_objects_Schlichting_Ofek_2012}, 
    although the precise slope index and the position of the break are not known exactly. 
    The break in the power-law index is expected because at lower radii
    the time-scale for collisional erosion is longer than the age 
    of the Solar System \citep{collisional_models_Dohnanyi_1969,KBOs_power_law_Pan_Sari_2004}.
    
    Measuring the number density at various sizes
    between 0.5 and 2\,km is useful for constraining
    the formation and collisional history of the Kuiper belt
    and has implications for understanding the conditions
    leading up to planet formation.
    The exact slope of the power law can help determine
    the material strength of small KBOs \citep{KBOs_power_law_Pan_Sari_2004}. 
    An overabundance of km-scale KBOs above a smooth power-law
    is expected by some models
    \citep{KBO_initial_sizes_Schlichting_2013}, 
    and the effects of such an excess 
    have presumably been observed in crater density measurements
    on large Kuiper belt objects such as Pluto and Arrokoth
    \citep{kuiper_belt_sizes_Pluto_craters_Morbidelli_2021}.  
    Measuring the size distribution of KBOs in this range
    will help confirm or rule out such models. 
            
    Km-scale KBOs cannot be detected directly
    by any existing observatories. 
    They can, however, be measured using 
    occultations of background stars. 
    Such occultations diminish a fraction of the star's light
    for $\sim 0.1$\,s 
    \citep{occultation_idea_Bailey_1976,KBO_occultation_diffraction_Roques_1987}.
    Detection of such occultations requires, in most cases, 
    short integrations and fast-readout cameras. 
    Two occultations by objects of $\approx 0.5$\,km 
    were detected using the Hubble Space Telescope 
    Fine Guidance Camera (HST-FGC; \citealt{KBO_single_object_Schlichting_Ofek_2009,abundance_kuiper_objects_Schlichting_Ofek_2012}). 
    Observing such events from the ground is more challenging, 
    due to atmospheric scintillations.
    Some dedicated surveys have been constructed 
    in hope of finding more KBO occultations
    \citep{TAOS_survey_Zhang_2013,TAOS_II_camera_Wang_2016,Colibri_survey_Pass_2018,OASES_survey_Arimatsu_2017}.
    So far, a single detection from a ground observatory has been reported, 
    of a 1.3\,km object \citep{KBO_detection_from_ground_Arimatsu_2019}. 
    
    The Weizmann Fast Astronomical Survey Telescope (W-FAST) 
    is a sub-meter class, wide-field telescope
    equipped with a fast camera 
    capable of observing a field of view of 7\,deg$^{2}$ 
    at a rate of 25\,Hz
    \citep{WFAST_system_overview_Nir_2021}. 
    
    Other surveys looking for KBO occultations have used a variety of
    methods to reduce such large photometric datasets. 
    For the data analysis for TAOS II, 
    \cite{occultation_real_time_detection_TAOS_II_Huang_2021} plan to use a 
    $\chi^2$ test on boxes of varying sizes to trigger each telescope 
    separately and all three telescopes combined. 
    While this method adapts to the expected width of the occultation pattern, 
    it necessarily loses some sensitivity due to the 
    difference between a box-shaped occultation light-curve 
    and the ingress/egress and diffraction fringes 
    expected for realistic occultations. 
	
    \cite{KBO_single_object_Schlichting_Ofek_2009, abundance_kuiper_objects_Schlichting_Ofek_2012} 
    used data from 
    the Hubble space telescope's fine-guidance sensor, 
    collected at 40\,Hz over several years, to find two KBO occultations. 
    Their method involved matching of the light-curves to occultation models. 
    They quantified the false positive rate using a bootstrap method
    of randomly sampling the flux measurements in time. 
    This method is useful for measuring the noise properties, 
    assuming there are no temporal correlations. 
    Since the data were collected 
    above the atmosphere, they do not suffer
    from most of the time correlations observed
    by ground-based photometric surveys. 

    For the data analysis for OASES, 
    \cite{KBO_detection_from_ground_Arimatsu_2019} used 
    the same rank statistics method applied by \cite{TAOS_analysis_Lehner_2010}
    in the first TAOS experiment. 
    This method looks for instances where the flux from multiple
    telescopes is relatively low at the same time. 
    To combine the significance of flux drops in multiple, consecutive frames, 
    they apply a smoothing filter with the expected occultation width
    before looking for simultaneous drops. 
    This method was used successfully to find a candidate KBO 
    occultation event in two telescopes' light-curve. 

    For the data analysis for the Colibri survey, 
    \cite{Colibri_survey_Mazur_2022} used a Ricker window\footnote{
        The Ricker window is the second derivative of the Gaussian, 
        $R(x)\propto (1-x^2/\sigma^2)\exp(-x^2/2\sigma^2)$. 
    } 
    to filter the stellar light-curves, and passed along any detections above 
    a threshold of 3.75$\sigma$ to be compared to more detailed KBO 
    occultation models, and cross correlated their timing between 
    the three telescopes in the observatory. 
    This method closely resembles the matched-filtering
    presented in this work, with the exception that the 
    initial filtering used the Ricker window and not 
    specific KBO models at the first step, 
    and assumed white (uncorrelated) noise. 
    Having multiple telescopes allows using 
    a much lower initial threshold than used here, 
    which compensates for the potential 
    loss of sensitivity due to the simplified filter shape. 
    
    In this work we present a matched-filter approach
    to detecting KBO occultations, 
    and its implementation for the W-FAST pipeline. 
    This includes pre-whitening of the frequency-dependent noise, 
    and injection of simulated occultations into real data 
    in order to estimate the detection efficiency of the algorithm. 
    We discuss the data quality cuts used to exclude bad data segments, 
    and present the MCMC procedures used to estimate the parameters
    for any occultation candidates found.     
    Some code in this work was adapted from \cite{matlab_package_Ofek_2014}. 

    While the analysis presented here focuses on the population of KBOs, 
    these methods can easily be adapted to searches for other trans-Neptunian populations, 
    although in those cases the typical sizes of the objects that would be detected
    will be larger, due to the larger diffraction length-scale, which scales 
    as the square root of the distance to the objects. 
    The durations of such occultations will also be longer, 
    such that they may be detectable using longer exposures than those discussed here. 
    For example, Oort cloud objects of sizes $\approx 10$\,km can be detected
    with durations of about half a second. 

    We present the input data in \S\ref{sec: data input}. 
    We discuss the occultation finding algorithm in \S\ref{sec: algorithm}, 
    the simulations used to model KBO occultations in \S\ref{sec: simulations}
    and the parameter estimation procedure in \S\ref{sec: parameter estimation}. 
    We conclude in \S\ref{sec: conclusions}. 
	
\section{Data input}\label{sec: data input}

    Although our algorithm is general, 
    its design centered around the W-FAST telescope \citep{WFAST_system_overview_Nir_2021}, 
    which is a wide field ($\approx 7$\,deg$^{2}$), 55\,cm telescope, 
    with a 4k$\times$4k sensor and a plate scale of 2.33''/pixel, 
    which regularly observes at a cadence of 10--25 frames per second.
    Throughout the paper we describe the analysis of the data obtained with W-FAST
    during the 2020--2021 observing periods. 
 
    The analysis is based on cutouts saved around bright stars.
    On disk, each file contains a batch of cutouts of 15 by 15 pixels, 
    with data for 100 images (frames) of a few hundred to a few thousand stars.\footnote{
    The W-FAST saves data products in HDF5 format, which has some advantages 
    in I/O speed as compared to FITS or ASCII files. 
    Each cutout file contains data for all detected stars, over 100 frames. 
    The first two dimensions (i.e., the memory continuous dimensions) 
    are for the $15\times 15$ pixel cutouts of each star, 
    the third dimension is for the frame number 
    and the last dimension is for the star index. 
    It should also be noted that the number of frames per file (100) 
    and the number of pixels per side for each cutout (15) are tunable 
    parameters and may change with future use of the W-FAST system. 
    }
    The stars are identified in the summed image of the first batch of 100 frames taken of each observing run on each field. A matched filter with a rough estimate of the system's point spread function (PSF) is used to suppress noise, 
    and any points above five times the remaining noise are marked as stars and are tracked throughout the run. 
    Since the summed image is deeper than individual images, 
    some of the identified stars would have a frame-to-frame photometric signal to noise (S/N) lower than 3, and would be excluded in the analysis process. 
    These files are loaded one by one and processed by
    a custom photometry pipeline described in \S\ref{sec: photometry}.
    Since the full dataset for a typical run of 1-2 hours 
    can reach more than 10\,GB, we only keep a subset of the data
    in RAM at any time. 
    The photometric products, i.e., the flux and other auxiliary measurements, 
    are stored in buffers in memory to provide some estimates of the 
    statistical properties of each data product. 
    At each point in the analysis, 
    we filter and process two adjacent batches, 
    with a total of 200 frames. 
    However, we only look for events with a flux measurement that surpasses the threshold 
    inside a 100 frame window in the middle of those 200 frames. 
    The 50 frames at the beginning and end of each 200 frame segment 
    are searched when processing the preceding or following batch. 
    This provides an overlap between batches, 
    and makes sure the search region is far enough away 
    from the edges of the data segment at any given time. 
    This helps when using cross correlations (matched-filtering) to avoid edge effects. 
    
    At the beginning of each analysis run, 
    we do not search for occultations until 
    having buffered at least 2000 frames. 
    This `burn-in' period is used to estimate the
    signal-to-noise ratio of each star, 
    and lets us disqualify a subset of the fainter and noisier stars
    before starting the occultation search. 
    It is also the minimal number of frames needed
    to estimate the Power Spectral Density (PSD) of each star. 
    
    The flux is first tested for sharp peaks caused mainly by cosmic rays. 
    Any flux value that is higher than eight times the local noise measurement 
    and immediately preceded and followed by values no larger than 3.5 times the noise, 
    is considered a cosmic ray event. 
    The value in that case is replaced using 
    piece-wise cubic spline interpolation. 
    
    If there are drifts in the stars' positions during an observations, 
    the cutouts are occasionally adjusted by a pixel in one or two directions. 
    This repositioning sometimes leads to changes in the mean flux of some of the stars
    due to the new position of the star's centroid relative to the photometric aperture. 
    When concatenating two batches of 100 images, 
    this sometimes causes jumps in the light-curve, 
    that can cause false detections. 
    To mitigate this we detrend the flux of each native batch, 
    before concatenating it with the previous batch. 
    Detrend in this context means fitting a linear function 
    to the flux values in each individual 100 frame batch, 
    and subtracting it from the original flux values. 
    These detrended fluxes are used for all subsequent analysis. 
    
    \section{Finding Occultations}\label{sec: algorithm}
    
    The pipeline is designed to maximize the 
    detection efficiency for sub-second occultations. 
    This is done by producing stable light-curves from the imaging data, 
    correcting these light-curves for low frequency noise, 
    and by matched-filtering the flux values with a template bank
    of simulated occultations that covers the parameter space of interest. 
    Whenever a part of one of the light-curves matches one of the templates
    we save that part of the data as an occultation candidate. 
    In what follows we discuss this pipeline in detail
    as well as the ways we disqualify sections of bad data
    and false alarm events. 
    Furthermore, in order to estimate the efficiency of the entire process
    we inject simulated events into real data and track
    which events are recovered by the pipeline.

    \subsection{Power Spectral Density correction}\label{sec: psd correction}

    The Power Spectral Density (PSD) is a measure of the noise power in each frequency bin. 
    The absolute value of the Fourier transform of a time series is a simple way to 
    estimate the power spectral density. 
    We use the backlog of the latest 2000 flux measurements
    to estimate the PSD for each star's flux.
    We use Welch's method~\citep{power_spectral_density_Welch_1967}
    of tapered, overlapping windows to calculate the PSD, 
    with a window of 100 frames and 50\% overlap. 
    This estimator is more resistant to momentary noise 
    incidents, that may bias the PSD values. 
    We update the PSD for each new batch of 100 frames using 
    a circular buffer of the latest 2000 frames. 
    We then use the PSD to correct the frequency dependent noise 
    in each light-curve. 
    Normally we would divide both the fluxes and the templates 
    by the square root of the PSD, i.e.,	
    \begin{equation}
        \widehat{f}_c(\nu) = \frac{\widehat{f}(\nu)}{\sqrt{\widehat{P}(\nu)}}, \quad \widehat{T}_c(\nu) = \frac{\widehat{T}(\nu)}{\sqrt{\widehat{P}(\nu)}},
    \end{equation}
    where $f$ are the detrended input fluxes, 
    $T$ is any of the templates used for matched-filtering, 
    $f_c$ and $T_c$ are the PSD corrected fluxes and templates, 
    $P$ is the PSD estimate, 
    and $\widehat{\square}$ is the Fourier transform from time coordinate $t$ 
    to frequency coordinate $\nu$. 
    Since we use the corrected template $T_c$ to multiply $f_c$, 
    we save computations by leaving the templates unchanged, 
    and dividing only the fluxes by the PSD:
    \begin{equation}
        \widehat{f}_c(\nu) = \frac{f(\nu)}{P(\nu)}.
    \end{equation}
    For more details, see \cite{gravitational_waves_non_gaussian_Zackay_2019}. 
    
    \subsection{Matched-filtering}\label{sec: matched filtering}
    
    Once the fluxes are corrected by the PSD, 
    we apply a matched-filter, calculating the filtered-fluxes:
    \begin{equation}
        f_f = T \star f_c,
    \end{equation}
    where $T$ is any template from the filter bank (see \S\ref{sec: template banks}), 
    $\square \star \square$ is the cross correlation operator,  
    which is just multiplication by the complex conjugate, 
    in Fourier space:
    \begin{equation}
        \widehat{f_f} = \widehat{T}\ \overline{\widehat{f_c}}, 
    \end{equation}
    where $\overline\square$ marks a complex conjugate. 
    We divide the filtered fluxes with a normalization term, $\sigma_f$, 
    which is estimated by calculating the RMS of the preceding 2000 measurements of $f_f$,
    for each star and each filter. 
    This ensures that the filtered fluxes are in units of $S/N$
    after the PSD correction. 
    
    The shapes we are trying to match are those 
    of KBO occultations. 
    We used custom simulation code to produce these templates, 
    as discussed in \S\ref{sec: simulations}. 
    
    We filter the data in two stages. 
    In the first stage we use a small template bank with only a few templates, 
    randomly selected to cover the parameter space 
    of possible occultations with up to 70\% overlap. 
    This means that any occultation in the parameter space 
    would recover at least 70\% of the $S/N$
    with at least one of the templates of the smaller bank
    (see \S\ref{sec: template banks}).     
    If at any time any star's flux, filtered with any of the templates in the small bank, 
    exceeds five times the noise estimate, 
    that star is flagged for further analysis. 
    Any stars that are flagged are then filtered by a larger bank
    with about $\times 10$ more templates, that has a minimal 95\% overlap with any template. 
    If those fluxes, filtered by any of the templates in the larger bank,
    exceeds 7.5 times the noise, 
    we save that star and that point in time as an occultation candidate. 
    
    The higher threshold of 7.5 times the noise 
    was chosen to deal with the large number of trials
    or flux measurements, 
    that can randomly produce high $S/N$ false events
    over the course of the survey. 
    A survey of 1000 hours, 
    with an average of 1000 stars, 
    taken at 25\,Hz, 
    will produce $\approx10^{11}$ independent flux measurements. 
    Even when assuming independent, normally distributed noise, 
    we would expect 0.1 or 0.003 false detections to occur 
    when setting a threshold $S/N$ of 7 or 7.5, respectively. 

    \subsection{Template banks}\label{sec: template banks}

    To generate template banks that have good coverage 
    without having to repeat similar templates, 
    we applied a random construction method 
    (e.g., \citealt{matched_filter_stochastic_template_placement_Harry_2009, 
    gravitational_waves_pipeline_Venumadhav_Zackay_2019, 
    gravitational_waves_non_gaussian_Zackay_2019}). 
    In this method we randomly generate templates 
    and check if they overlap with any of the templates already in the bank:
    \begin{equation} \label{eq:overlap}
        O = \max(T_i \star T_j), \text{ where } \sqrt{\Sigma_t T_i^2(t)} = 1. 
    \end{equation}
    The sum over $t$ normalizes the templates such that a template cross-correlated with 
    itself would give an overlap of $O=1$. 
    If a new template does not overlap enough, 
    i.e., the maximum $S/N$ it would get when detected by any of the existing templates is too low, 
    then it is added to the bank. 
    If the overlap is higher than a given threshold, it is not added. 
    For the smaller template banks we require $O>0.7$ and for the full bank $O>0.95$. 

    Each new template's parameters are chosen uniformly from the range of allowed parameters: 
    stellar sizes of 0--3 Fresnel Scale Units (FSU), 
    occulter radii between 0.3--2\,FSU, 
    impact parameter of 0--2 FSU, 
    and velocities of 5--30\,FSU\,s$^{-1}$. 
    The Fresnel unit is equivalent to $\sqrt{\lambda D /2}$, 
    which, at a central wavelength $\lambda=550$\,nm and a KBO distance of $D=40$\,AU 
    is approximately 1.3\,km.     
    The stellar size is the angular size of the background star, 
    converted to Fresnel units projected to the distance of the occulter. 
        
    We continue trying new templates until $2\times 10^4$ are rejected, 
    at which point we save the template bank to be used in the full analysis. 
    For the small template banks we use an overlap threshold of 0.7, 
    and for the full template banks we use an overlap threshold of 0.95. 
    This means that the full (small) template bank should trigger with at least
    95\% (70\%) $S/N$ to any signal within the parameter space.     
    For data taken at 25\,Hz we use 32 templates in the smaller bank, and 307 in the full bank. 
    For data taken at 10\,Hz we use 18 templates in the smaller bank, and 507 in the full bank. 
    The exact number of templates depends on the range of parameters, 
    the specific wavelength and frame rate used, and the random instances 
    chosen when generating the template bank, 
    so would be different for each survey's pipeline. 

    The number of templates we find necessary to cover 
    the parameter space is substantially larger than one, 
    which implies that surveys using a small number of templates, 
    or even a single template, may underestimate the occultation rate of KBOs by a factor of a few.

    \subsection{Star hours} \label{sec: star hour calculation}
    
    To estimate the rate of occultations it is necessary to know
    how many hours of usable observations we have accumulated, 
    multiplied by the number of stars, as a function of the star's $S/N$,
    hence: star-hours. 
    Each batch of data that is scanned, specifically, 
    the middle 100 frames we use to search for candidates, 
    can contribute at maximum four seconds of useful time 
    (or ten seconds at 10\,Hz)
    for each star we are tracking. 
    This number is often lower than this upper limit, 
    since some of the stars have $S/N$ too low to use for event detection 
    (stars with $S/N<3$ per measurement are not used), 
    and some sections of the data do not pass our data quality cuts
    (as discussed in \S\ref{sec: quality cuts}). 
    
    Furthermore, not all star-hours are equally useful.
    Stars with high $S/N$ can be used to detect shallower occultation, 
    e.g., by smaller occulters or with a higher impact parameter. 
    Thus, we accumulate star hours in bins of $S/N$, 
    with a bin width of 0.5, spanning the range between 2 and 40, 
    which are the typical values for the photometric $S/N$ we measured at 10--25\,Hz. 
    
\begin{figure}
        
        \centering
        
        \pic[1]{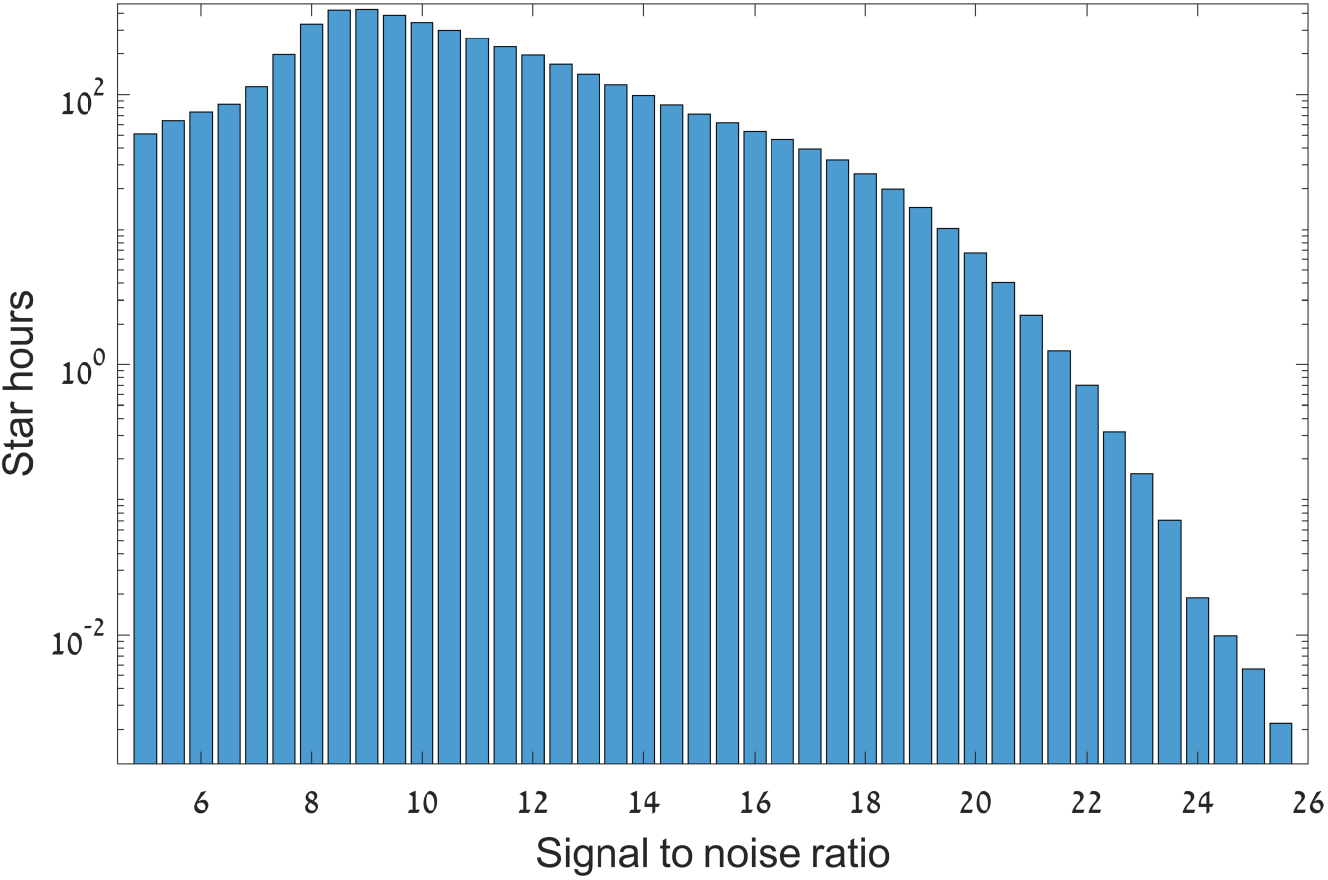}
        
        \caption{An example for star hours in different star $S/N$ for a run 
                 taken on 2020 June 9th,  
                 centered on RA$=19^{h}37^{m}25.6^{s}$, Dec$=+38^\circ56'50.6''$ (J2000). 
                 The total duration of the observations was 1.27 hours, 
                 photometrically following 2000 stars in the field.
                 The sum of useful hours for all stars accumulated to 2332.6 star-hours, 
                 with losses due to quality cuts reducing the total to 2125.2 star-hours
                 (a loss of 207.4 star-hours). 
            }
        \label{fig: example run star hours}
        
    \end{figure}

    An example histogram of star-hours for different $S/N$ values 
    is shown in Figure~\ref{fig: example run star hours}. 
    This shows the star-hours we gained from observing a dense field
    at RA$=19^{h}37^{m}25.6^{s}$, DEC$=+38^\circ56'50.6''$ (J2000), 
    which is an off-ecliptic field 
    which was used as control observations for the KBO search.
    We took a total of 1.27 hours of continuous data in this run 
    which began on UTC 2020 June 9th, 20:17:37. 
    In this case we extracted photometry for 2000 stars 
    with Gaia magnitudes of $8 \lesssim B_p \lesssim 12.5$, 
    at a frame rate of 25\,Hz. 
    The star hours do not include the times rejected by any of the quality cuts. 
    
    \subsection{Data quality cuts} \label{sec: quality cuts}
    
    In parallel to event finding, 
    we also run a data-quality pipeline 
    that continuously checks the raw flux and auxiliary data
    for indications that the light-curve quality is poor,
    e.g., when the centroid is far from the center of the cutout, 
    or if there are correlations between the flux and the background measurements, 
    indicating, in this case, leakage of starlight from the star or one of its neighbors 
    on to the photometric annulus. 		
    In some cases the cut disqualifies the entire batch, 
    but in most cases only some frames or some stars are affected. 
    The quality cuts made on the data are listed in \S\ref{sec: quality cuts list}.
    
    Each one of these cuts is used to flag some of the frames, 
    so that any occultation candidates that are detected
    during the flagged times and on the	flagged stars
    can be disqualified and not treated as real occultations. 
    Since we would not want to use times that are too close to a disqualified frame, 
    we also disqualify ten frames before and after any frame that failed any cut. 
    This increases the number of flagged frames/stars, 
    but also helps disqualify event candidates where the peak is nearby, 
    but not overlapping, a region of bad data. 
    
    For each batch of analyzed data, 
    we keep track of the total amount of star-hours that was searchable
    (as described in \S\ref{sec: star hour calculation}). 
    The excluded regions, after expanding them by ten frames on either side, 
    are not counted as usable times. 
    Those times are counted for each different cut as `time losses', 
    giving us a sense of the coverage lost due to each cut. 
    The number of star hours lost is shown in 
    Table~\ref{tab: cut losses}. 
    We present both exclusive and inclusive losses. 
    Exclusive losses are times 
    that were lost only to the specific cut, 
    which is the amount of time 
    that would be recovered if this cut was removed. 
    Inclusive losses are times lost to the specified cut
    regardless of overlap with other cuts, 
    which is the amount of time
    that would be lost even if all other cuts were removed. 
    
    \begin{table*}
		
    \centering
    \caption{The number of star hours lost to various quality cuts, 
        which are described in \S\ref{sec: quality cuts list}. 
        Exclusive losses are those times lost only to the specified cut. 
        Inclusive losses are those times lost to a cut regardless of other cuts. 
        Bad stars losses include all star hours recorded for stars that 
        had more than five triggers during the same run. 
        The total number of star hours were collected over 2020--2021, 
        including all the relevant observations at high cadence (10--25\,Hz).
        }
    
    \begin{tabular}{l|c|c}
    Cut name                     &   inclusive[h]    &   exclusive[h]    \\
        \hline \hline
        \code{delta_t}               & 40946.95 ( 2.33\%) & 10030.75 ( 0.57\%) \\
        \code{shakes}                &  1380.80 ( 0.08\%) &   440.03 ( 0.03\%) \\
        \code{fwhm}                  & 126313.11 ( 7.19\%) & 53937.32 ( 3.07\%) \\
        \code{instability}           &  3776.05 ( 0.21\%) &   324.27 ( 0.02\%) \\
        \code{slope}                 &     0.95 ( 0.00\%) &     0.00 ( 0.00\%) \\
        \code{near_bad_rows_cols}    & 20098.60 ( 1.14\%) &  6335.99 ( 0.36\%) \\
        \code{offset_size}           &  2948.82 ( 0.17\%) &   116.71 ( 0.01\%) \\
        \code{mean_flux_norm}        & 148555.33 ( 8.46\%) & 101455.34 ( 5.78\%) \\
        \code{linear_motion}         & 37207.05 ( 2.12\%) & 19251.18 ( 1.10\%) \\
        \code{background_intensity}  & 70200.95 ( 4.00\%) & 27469.41 ( 1.56\%) \\
        \code{aperture_difference}   & 81610.38 ( 4.65\%) & 17343.37 ( 0.99\%) \\
        \code{bad_pixels}            & 34456.45 ( 1.96\%) & 15095.42 ( 0.86\%) \\
        \code{repeating_columns}     & 43870.18 ( 2.50\%) & 12194.34 ( 0.69\%) \\
        \code{rolling_rms_25}        &     0.00 ( 0.00\%) &     0.00 ( 0.00\%) \\
        \code{rolling_rms_50}        &     0.00 ( 0.00\%) &     0.00 ( 0.00\%) \\
        \code{flux_corr_25}          & 15529.89 ( 0.88\%) &     3.16 ( 0.00\%) \\
        \code{flux_corr_50}          & 80967.72 ( 4.61\%) & 29045.31 ( 1.65\%) \\
        \code{corr_b_25}             &  1722.41 ( 0.10\%) &   497.27 ( 0.03\%) \\
        \code{corr_b_50}             &  3529.45 ( 0.20\%) &  1147.64 ( 0.07\%) \\
        \code{corr_x_25}             & 13813.63 ( 0.79\%) &  1256.26 ( 0.07\%) \\
        \code{corr_x_50}             & 42045.56 ( 2.39\%) &  5918.31 ( 0.34\%) \\
        \code{corr_y_25}             & 12944.24 ( 0.74\%) &  1231.20 ( 0.07\%) \\
        \code{corr_y_50}             & 36636.43 ( 2.09\%) &  5034.37 ( 0.29\%) \\
        \code{corr_r_25}             & 45153.80 ( 2.57\%) &  3493.14 ( 0.20\%) \\
        \code{corr_r_50}             & 176160.75 (10.03\%) & 70104.28 ( 3.99\%) \\
        \code{corr_w_25}             & 28419.65 ( 1.62\%) &  1552.36 ( 0.09\%) \\
        \code{corr_w_50}             & 76973.47 ( 4.38\%) & 23401.58 ( 1.33\%) \\
        \code{Bad stars}             &    97.80 ( 0.01\%) &         ---       \\
        \hline
        good/total star hours        & 1070837.4h (60.97\%) out of 1756332.6h

    \end{tabular}

    \label{tab: cut losses}
		
    \end{table*}
	
    Event detection is kept separate, as much as possible, 
    from the quality cuts made on the data. 
    After occultation candidates are detected, 
    we compare the time of event peak $S/N$ with the various quality cuts. 
    If that frame overlaps with any bad data frames 
    (or if it is ten frames before or after it)
    the event is disqualified. 
    
    On top of these quality cuts, 
    we also disqualify batches or stars that have more than five 
    candidate detections (regardless of whether these candidates passed the other cuts). 
    Since true occultation events are rare 
    (on the order of one in $10^4$--$10^5$ star-hours), 
    we disqualify any star that produces five or more candidates in one observing run 
    (typically one or two hours) or a batch with five or more candidates. 
    In most cases this disqualifies a small subset of stars that are 
    positioned near another star or a sensor artefact, 
    or particularly noisy batches where the telescope 
    was affected by, e.g., wind gusts. 
    
    Finally, if a certain run has more than ten consecutive batches 
    that are ``bad'', i.e., have a background $>50$ counts pixel$^{-1}$
    or star widths $>20''$, 
    the run is aborted mid-way. 
    This saves some analysis time for runs taken as the sun is rising, 
    or when the telescope is badly de-focused.

    \subsection{Injection of simulated light-curves}\label{sec: injection simulations}
	
    We inject simulated events into observed light-curves, 
    which are used to quantify the efficiency 
    of the detection algorithm and human vetting (see \S\ref{sec: human vetting}). 
    Each batch of 100 frames (4 or 10 seconds) has a 25\% probability 
    to have a simulated event injected into its flux data. 
    This is done after regular event detection is completed for that batch. 
    A star is picked at random from stars that did not pass the smaller (70\% overlap) template bank. 
    This makes sure that the star does not have a pre-existing occultation-like light-curve. 
    
    The flux from two consecutive batches is used as the baseline for each injected event, 
    where the peak of the simulated event is centered on a randomly chosen time frame out of the middle 100 frames
    (the region searched by the pipeline). 
    We begin by taking the detrended flux
    as used by the regular analysis:
    \begin{equation}
        f_d = f_r - P_1(f_r),
    \end{equation}
    where $f_r$ and $f_d$ are the raw and detrended fluxes in each batch, 
    and $P_N(f)$ is the $N$-th order polynomial fit to the flux $f$. 
    We then extract any residual slow variations in the light-curves, 
    by fitting a 2nd order polynomial to the detrended flux over the entire 200 images. 
    The difference between the de-trended flux and the polynomial fit
    represents the noise for that event:
    \begin{equation}
        f_n = f_d - P_2(f_d).
    \end{equation}
    We calculate the mean of the (background subtracted but not detrended) flux over 200 frames, 
    which represents the baseline flux of the star. 
    We add this mean flux to the 2nd order polynomial we calculated earlier: 
    \begin{equation}
        f_s = P_2(f_d) + \langle{f_r - f_b}\rangle,
    \end{equation}
    where $f_b$ is the background flux 
    and $\langle\square\rangle$ represents an average over 200 frames. 
    The smoothed flux $f_s$ represents the ``noise-less'' flux from the star, 
    upon which we can superimpose the simulated occultation. 
    
    Then we choose a random simulated occultation light-curve $T_\text{sim}$ 
    from the range of allowed occultation parameters. 
    The template is shifted so that the mid-point of the occultation is placed 
    randomly in the search region (the 100 frame region in the middle of the detrended 200 frames)
    while avoiding frames that were disqualified by quality cuts. 
    
    We then apply the template $T_\text{sim}$ to the smoothed flux:
    \begin{equation}
        f_{s,\text{sim}} = T_\text{sim} f_s,
    \end{equation}
    and re-scale the noise based on the relative variance:
    \begin{equation}
        f_{n,\text{sim}} = f_n \sqrt{\frac{T_\text{sim}V_f + V_b}{V_f+V_b}},
    \end{equation}
    which scales the noise down in areas where the occultation 
    reduces the flux and thus the source noise in that time range. 
    We assumed a noise model of
    \begin{equation}
        V_\text{sim} = V_f + V_b,
    \end{equation}
    where $V_f$ is the source flux variance, assuming Poisson statistics\footnote{
        We find the brightest stars tend to have non-Poisson noise, 
        that scales linearly with the flux (instead of the square root of the flux). 
        This includes only a small fraction of the stars, so this assumption is reasonable. 
    },
    and $V_b$ is the background variance calculated for that light-curve
    using the annulus around the star. 
        
    The simulated occultation template $T_\text{sim}$ is produced by the same 
    tools we use for producing filter templates. 
    The light-curve $f_\text{sim}$ thus has dips
    that are proportional to the depth of the occultation and
    the mean flux of the star. 
    Finally, we add the background noise that was subtracted
    in the initial phase, 
    so that the light-curve is equivalent to 
    real measured light-curves from stars. 
    The de-trended flux is calculated by adding 
    the injected smoothed flux to the adjusted noise: 
    \begin{equation}
        f_{d,\text{sim}} = f_{s,\text{sim}} + f_{n,\text{sim}}.
    \end{equation}
    The injected, detrended light-curves go through the same detection pipeline, 
    with the exception that they directly go to the full template bank.\footnote{
        The smaller template bank is less restrictive than the full sized bank, for occultation events, 
        and is used in the analysis only to save run time. 
        For simulated events, the correct star is known \emph{a priori} so there is no need
        to pre-filter many light-curves using the smaller bank. 
        We assume any event that passes the higher threshold of the full template bank would have also 
        passed the low threshold of the small bank. 
        This was tested on simulated (white) noise and shown to be true for $>99$\% of events. 
    }    
    Any candidate events that pass the threshold are saved along with the real events. 
    They are flagged as simulations, 
    but this flag is not exposed to the human scanner until the analysis is done
    (see \S\ref{sec: human vetting}). 
    
    The simulated events have randomly chosen input parameters. 
    The four key parameters are 
    stellar size, $R_\star$, 
    occulter radius, $r$, 
    impact parameter, $b$, 
    and transverse velocity, $v$. 
    We choose $R_\star$ based on the actual 
    star radius, which we fit using the Gaia effective temperature and magnitude, 
    that is extrapolated to a bolometric magnitude using the Gaia color 
    (using the Gaia DR2; \citealt{GAIA_data_release_two_2018}). 
    If the star where we inject the simulation does not have a solved radius
    (because it did not have a Gaia fit or if there was a problem calculating its bolometric magnitude)
    then we estimate its size based on its $S/N$, ignoring its color, 
    by using a 2nd order fit of stellar size to $S/N$ which is calculated 
    at the beginning of each run based on the stars that do have Gaia fits. 
    The velocity and impact parameter are chosen from uniform distributions 
    of $0\leq b \leq 2$\,FSU and $5 \leq v \leq 30$\,FSU\,s$^{-1}$. 
    The occulter radius is chosen from a power-law with an index of $q=3.5$. 
    A steep power-law index is chosen to allow more small occulters 
    to trigger the detection algorithm, otherwise the detection
    numbers in the low-radius range would be too small. 
    The value is similar to that of \cite{abundance_kuiper_objects_Schlichting_Ofek_2012}, 
    but the exact choice is not so crucial, 
    as the results of the efficiency calculations
    are usually calculated in separate radius bins. 
    
    Because the majority of simulated events have a small radius, 
    only a small fraction (on the order of 10\%) of the simulated events
    ever pass the detection threshold. 
    Thus we are left with a few simulated candidate events
    for every 100 batches (10,000 frames). 
    A typical run with 1000--2000 batches would contain 
    about 10--20 simulated candidates, 
    and a handful of false positives due to 
    tracking errors and other artefacts.

    \subsection{Human vetting of candidates}\label{sec: human vetting}
    
\begin{figure}
        
        \centering
        
        \pic[1]{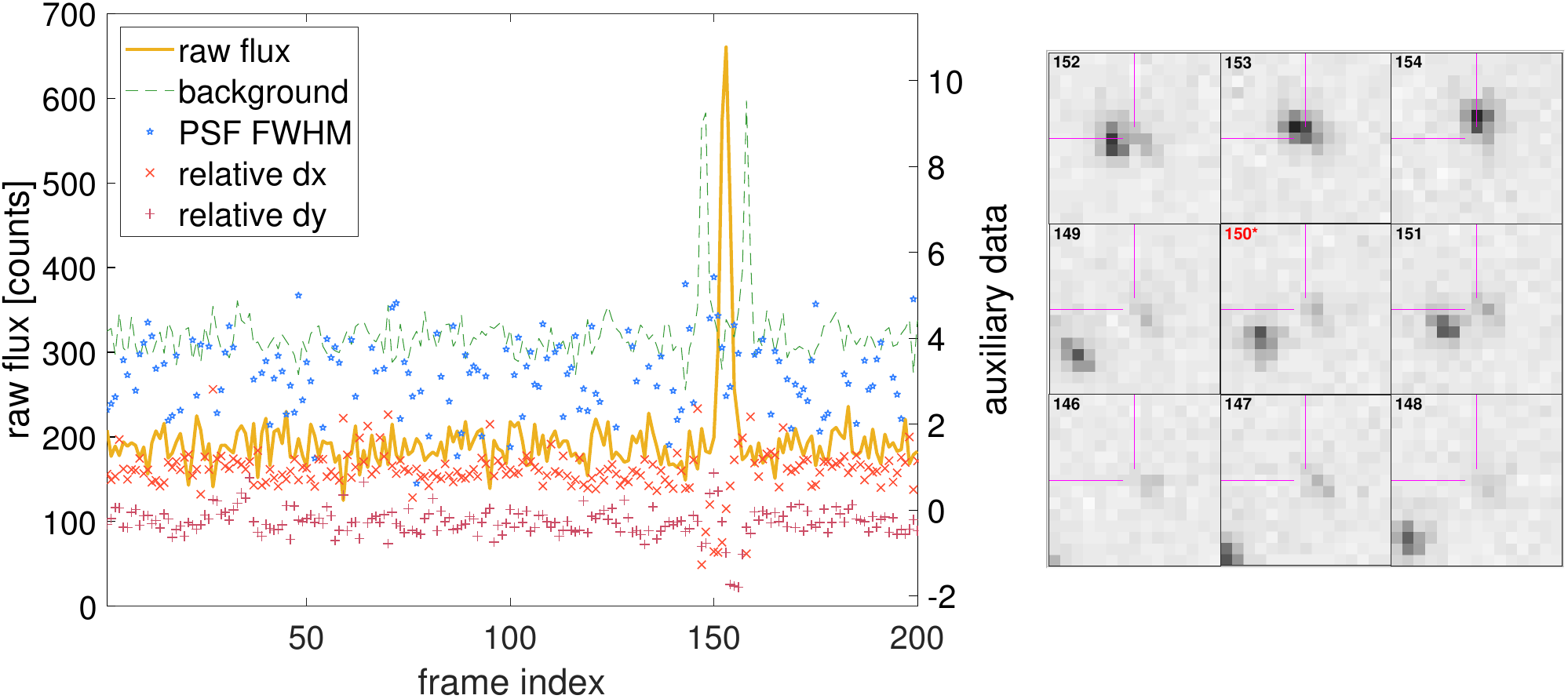}
        \caption{Light-curves (left) and cutouts (right) for a flare caused by a satellite passing near a star.}
        \label{fig: example satellite flare}
        
        
    \end{figure}

    Each saved candidate, including simulated events, 
    is processed visually by a human scanner. 
    This is useful as some rare events are clearly not astrophysical, 
    but are also not excluded by the various quality cuts. 
    The scanner looks at each candidate, 
    reviewing the light-curve of the star, 
    the cutouts taken around the time of the event, 
    and other auxiliary data. 
    Based on these data, 
    the scanner can decide what the most likely
    classification for the event would be, 
    either a clear occultation, 
    a possible occultation where the scanner is unsure, 
    or other categories such as satellite, cosmic ray, etc. 
    In most cases the occultation candidates are simulated events
    that are injected into the real data (see \S\ref{sec: injection simulations}), 
    and the non-occultation candidates are clearly non-astrophysical phenomena. 
    The scanner does not know which event is real and which is simulated 
    when making the classification decisions. 
    An example for such an event is shown
    in Figure~\ref{fig: example satellite flare}, 
    where a moving source (a satellite) is clearly seen.\footnote{
        In other cases, stars are shown to brighten dramatically
        for a few frames, without any apparent motion of a bright source
        on top of the star. 
        While it is possible that some of these events are actually
        positive-occultations, as explained in \S\ref{sec: binary occulters}, 
        we cannot rule out that they are glints from geosynchronous satellites
        (e.g., \citealt{satellite_glints_Nir_2021})
        and so we do not classify them as occultations. 
    } 
        
\begin{figure}
        
        \centering
        
        \pic[1]{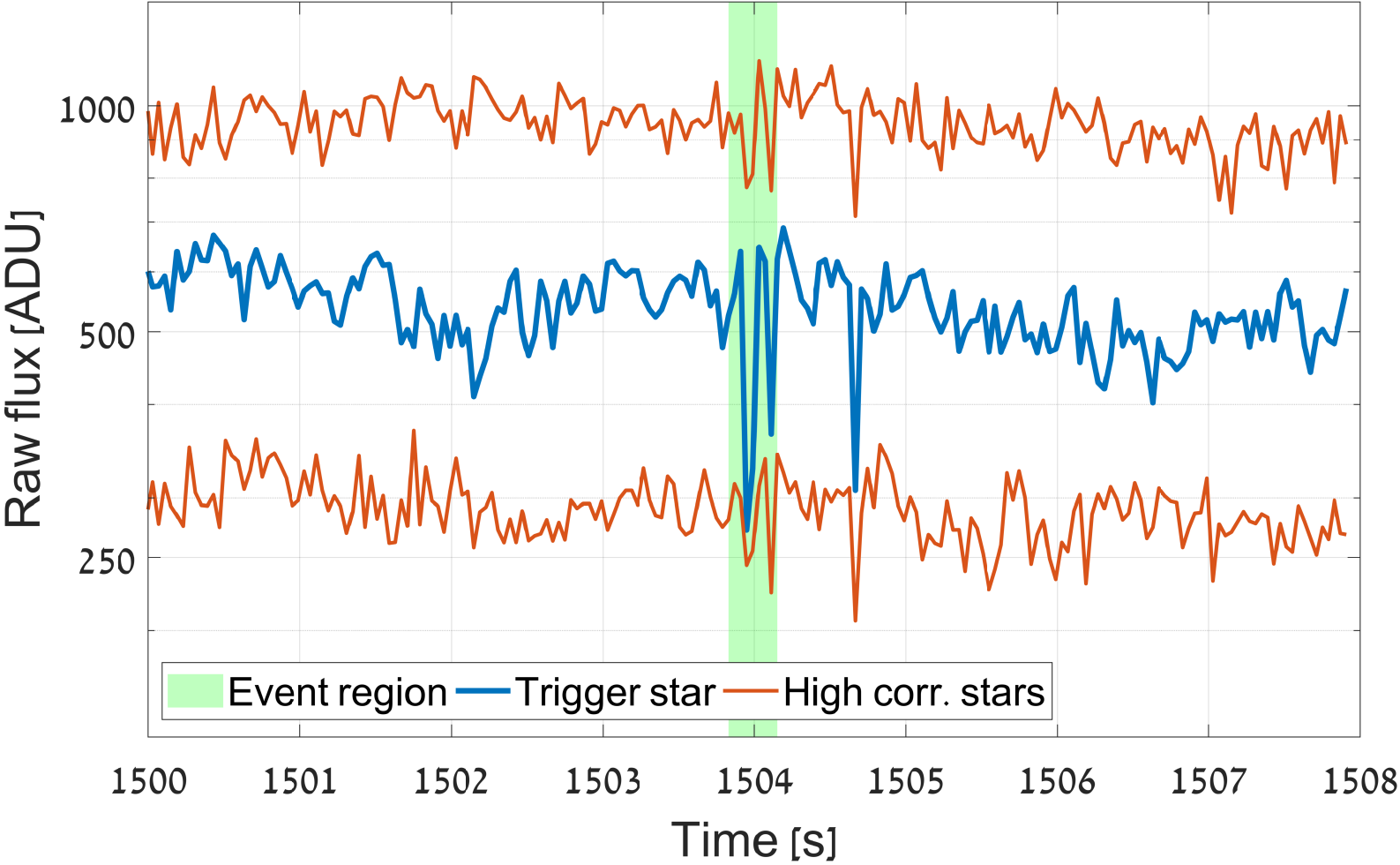}
        \caption{Light-curves for several stars, showing strong correlations around the triggered event time. 
                 }
        \label{fig: tracking error event}
        
        
    \end{figure}

    Another type of false positive is due to tracking errors or localized clouds, 
    which affect several stars at the same time. 
    Such events are identified by finding flux correlations 
    between the trigger star's flux and other stars in the field. 
    We find the three stars that have the highest correlation 
    between their flux and the trigger star's flux, 
    and present the light-curves of those stars. 
    The human scanner can check if the light-curves 
    show similar behaviour around the time of the trigger.
    We show an example for such correlated light-curves 
    in Figure~\ref{fig: tracking error event}.
    Some of these tracking error triggers are
    excluded by the \code{flux_corr} quality cuts, 
    but occasionally a few will pass and will have
    to be disqualified by the human scanner. 
    Satellites and tracking errors are the two most significant
    sources of false detections, and are usually easy 
    to tell apart from occultation events. 
    	
\begin{figure}
        
        \centering
        
        \pic[1]{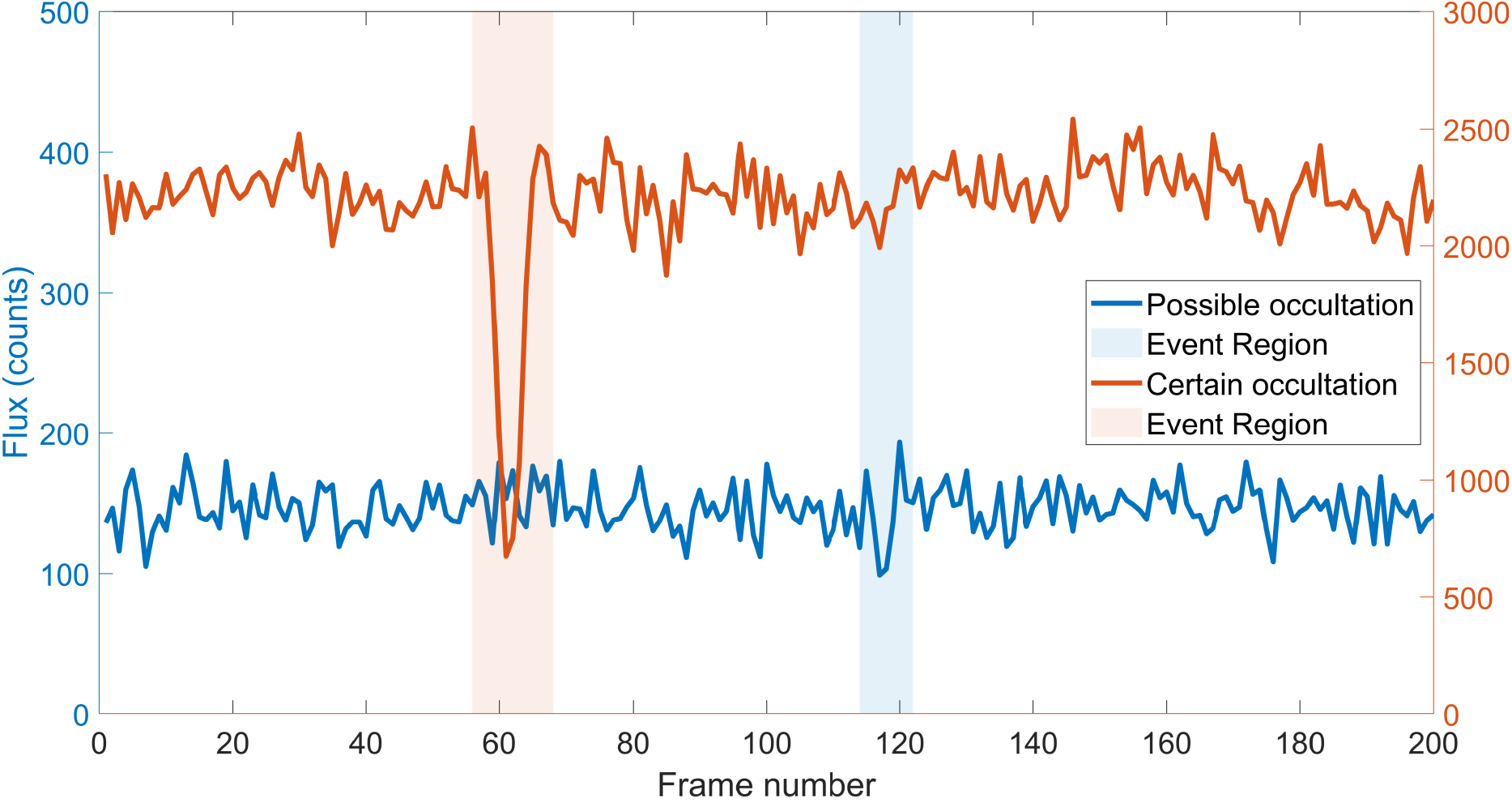}
        \caption{Light-curves for simulated events classified as 
                 a certain- and a possible-occultation. 
                 For each light-curve the event region 
                 is highlighted by a shaded region of 
                 the appropriate color. 
                 For the certain occultation (top, red curve), 
                 the drop in the light-curve is deep, 
                 due to the large radius of the occulter
                 ($r=1.64$ Fresnel units). 
                 Such events are easily detected and classified. 
                 However, due to the steep power-law index of the size distribution, 
                 they are expected to be rare compared to shallower events. 
                 For the possible occultation (bottom, blue curve), 
                 the drop is only slightly more pronounced
                 than the surrounding noise fluctuations. 
                 In other cases, an occultation could be 
                 classified as ``possible'' if other stars'
                 light-curves show some correlations with 
                 the event shape (but to a lesser degree than
                 the correlations shown in Figure~\ref{fig: tracking error event}), 
                 if the star is de-focused, 
                 or if the photometric $S/N$ is very low, etc. 
                 }
        \label{fig: example occultations}
        
        
    \end{figure}

    If a candidate does not appear to be  
    any kind of false positive, 
    the scanner assesses the shape 
    of the light-curve and classifies the 
    event as either a certain occultation
    or a possible occultation. 
    Since the vast majority of such candidates
    are simulated events, 
    we can estimate how many events would be 
    accepted or rejected by the pipeline and human scanner. 
    Generally the `certain' occultations are
    those with a large occulter radius, 
    and cause large, clear dips in the light-curve. 
    If such an event occurs in our data, 
    the human scanner will very likely not miss it.
    An example for such a simulated event is shown 
    in Figure~\ref{fig: example occultations} (top, red curve). 
    The `possible' occultations are those with a
    lower radius, or a higher impact parameter. 
    An example for such an event is shown in 
    Figure~\ref{fig: example occultations} (bottom, blue curve).
    
    It is important to note that the scanner
    is not made aware if a candidate is simulated. 
    Also, we do not reveal the coordinates of the field, 
    if it is on the ecliptic or not, 
    or what the Earth's projected velocity is on that field. 
    The projected velocity dictates the width of the occultations, 
    and can be used to disqualify events that are too wide or narrow. 
    By blinding the scanner to this information, 
    we remove some of the bias, e.g., 
    a higher tendency to accept candidates
    when scanning observations on the ecliptic plane. 
    Later, after some non-simulated events are classified as occultations, 
    we can compare their best-fit velocity to the Earth's projected velocity, 
    and see if the detections have a consistent velocity.

    \subsection{Estimating detection efficiency}\label{sec: detection efficiency}
    
    With extensive injection simulations we can estimate what
    fraction of occultation events would be detected by our pipeline. 
    This includes the full analysis pipeline from the raw flux stage, 
    up to the human vetting of candidates. 
    Here we present an example for our detection efficiency 
    based on 6220 events injected into randomly chosen stars
    in images recorded during 2020 July 1--18. 
    We use this simulated event sample, injected into real data, 
    as the base for the following results. 
    Real data has correlated noise, 
    so the probability to measure values in one measurement 
    is not independent of the measurements before and after it. 
    This reduces the sensitivity of a matched-filter trigger algorithm, 
    since the shape of the event that is matched needs to stand out 
    of noise that itself has random structure over different time scales. 
    
\begin{figure}
        
        \centering
        
        \pic[0.8]{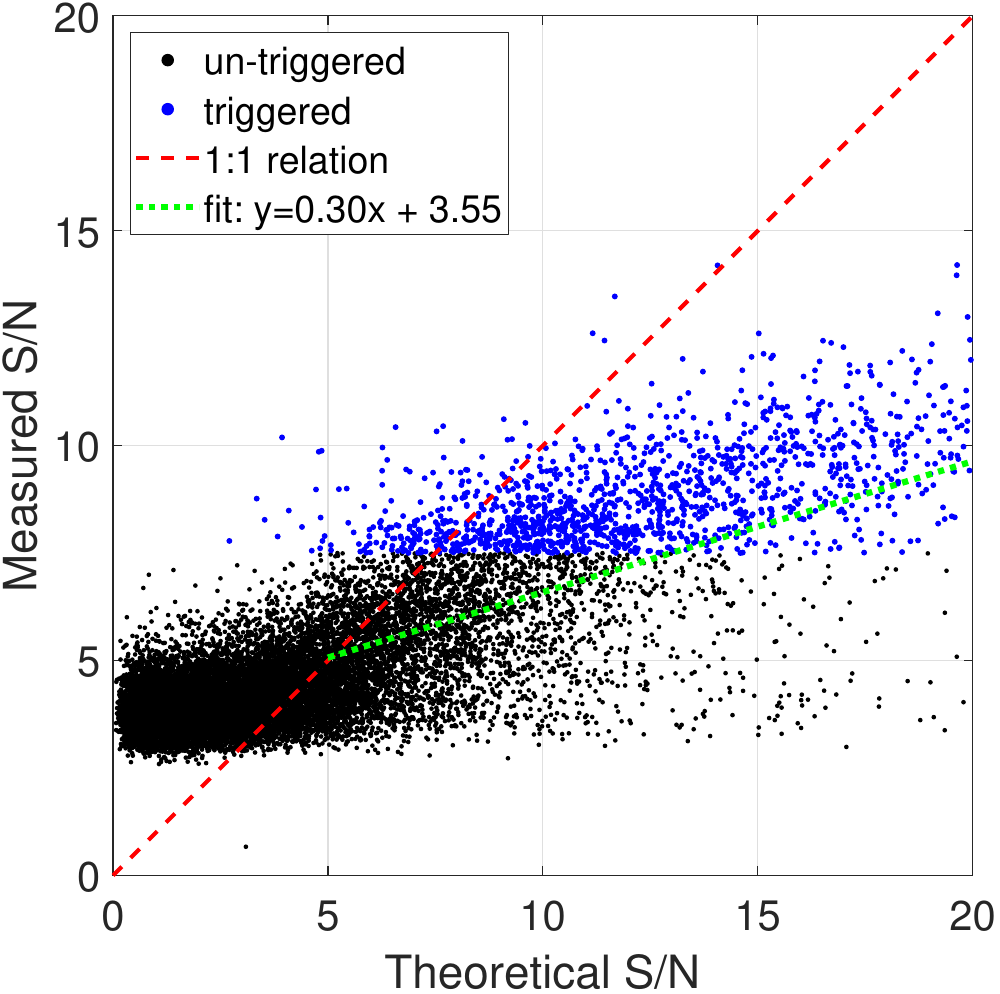}
        \caption{Measured vs.~theoretical $S/N$ for simulated events injected 
                 into real, correlated noise light-curves. 
                 The blue and black points represent individual events, 
                 triggered and un-triggered, respectively. 
                 The red line represents the 1:1 relation, 
                 whereas most of the events with $S/N>5$ lie below this line. 
                 The low $S/N$ events cluster around values of 3--5, 
                 which are presumably due to random noise, 
                 and not to the signal from the injected templates. 
                 The green dotted line is a linear fit to the theoretical $S/N>5$ points, 
                 showing that only 30\% of the $S/N$ is recovered in these data, 
                 presumably due to correlated noise. 
                 This demonstrates the importance of using 
                 a PSD correction: 
                 without it, the measured $S/N$ could be as much as $\times 3$ 
                 bigger than a truly representative score. 
              }
        \label{fig: snr detect vs theoretical}
        
                
    \end{figure}

    We check if the detection $S/N$ we find in injections into actual data 
    is consistent with the theoretical $S/N$ that we calculate
    based on the shape of the occultation template and assuming normal i.i.d.~noise
    with amplitude consistent with the chosen star's photometric noise. 
    For triggered events we use the highest $S/N$, which is recorded by the pipeline.
    For un-triggered events, such that did not pass the pipeline threshold, 
    we record the highest $S/N$ for that batch, in any template. 
    The results are presented in Figure~\ref{fig: snr detect vs theoretical}. 
    The points show the value of the measured $S/N$ as a function of the theoretical $S/N$,
    for triggered (blue) and un-triggered events (black). 
    The injection events do not follow the 1:1 relation (dashed, red line)
    for two reasons: 
    (a) at low values of $S/N\lesssim 5$ the templates give almost no signal, and the pipeline records 
        the highest score achieved by random noise in that batch, with $S/N\approx$ 3--5;
    (b) at high values of $S/N\gtrsim 5$ the correlated noise in the real data
        degrades the $S/N$ and reduces the detection efficiency. 
    A linear fit to the data (triggered and un-triggered) with theoretical $S/N>5$ is shown as a green, dotted line, 
    where the slope of the curve shows that only 30\% of the theoretical $S/N$ is recovered 
    from the correlated noise of real light-curves. 
    While the PSD correction does give optimal weights for different frequencies, 
    as explained in \S\ref{sec: psd correction},
    it also degrades some of the signal. 
    This is an inescapable loss of information 
    due to the higher noise levels at low frequencies, 
    and highlights the importance 
    of estimating the detection efficiency
    using real data, rather than theoretical estimates
    based on white noise. 
    Specifically, ground surveys that do not account
    for red-noise in the data
    will suffer from considerable loss of efficiency, 
    and possibly inaccurate rate estimates. 
    
\begin{figure*}
        
        \centering
        
        \pic[1]{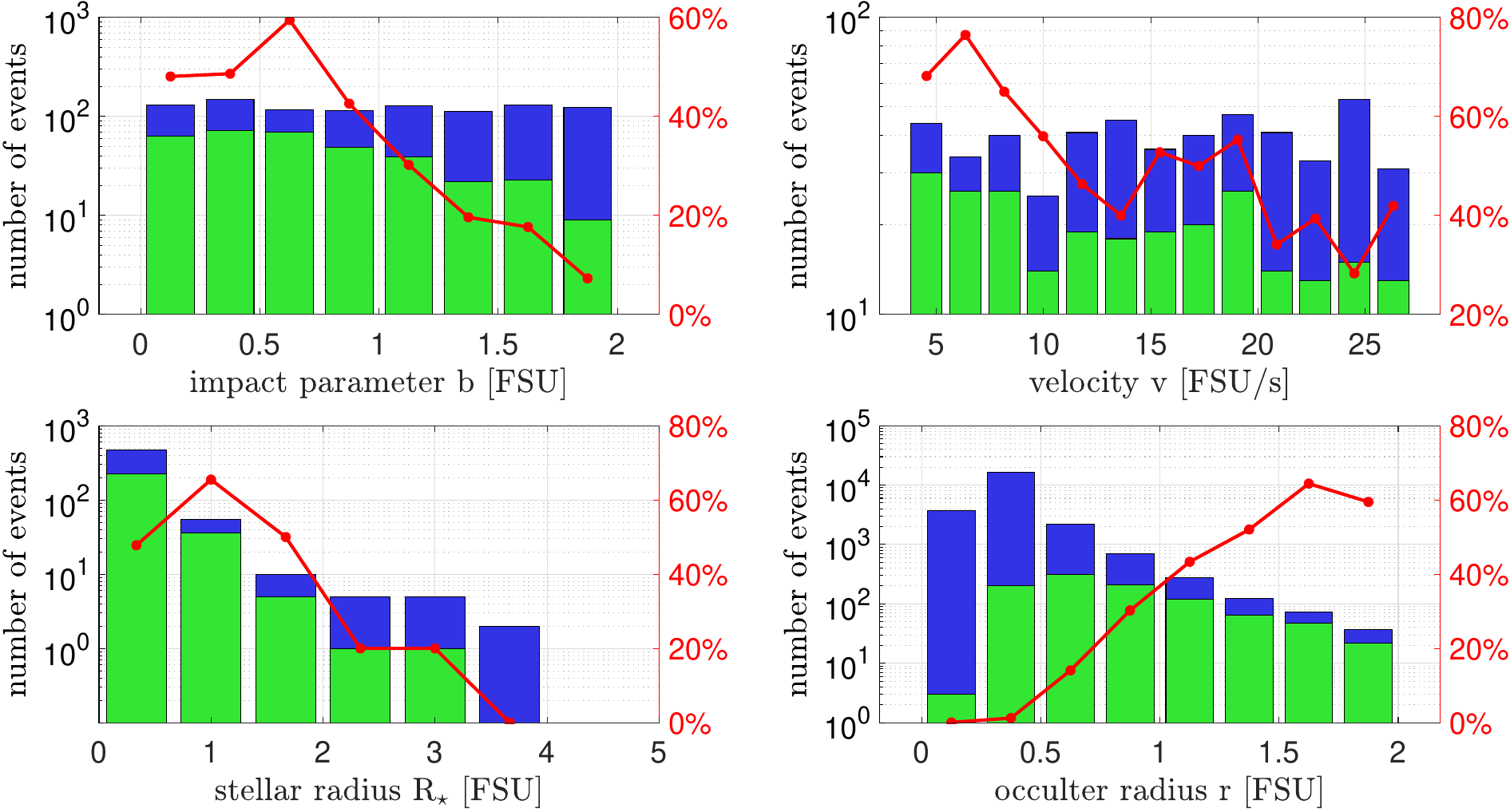}
        \caption{The number of simulated, detected events (green bars) 
                 out of the total injected events (blue bars)
                 for various occultation parameters. 
                 To increase the detection ratio and better explore 
                 the dependency on each parameter separately,
                 we applied the following pre-selection criteria for the events: 
                 $r\geq 1$\,FSU, $R_\star\leq 1$\,FSU, $b\leq 1$\,FSU, 
                 and $v\leq 20$\,FSU\,s$^{-1}$, 
                 where each panel used the limits on the three parameters
                 not explored in that panel. 
                 For example, the bottom right panel explores the effects of the occulter radius $r$, 
                 so the events used in that plot are pre-selected using $R_\star$, $b$ and $v$. 
                 Since all panels except that one use $r>1$\,FSU, they all display much smaller event numbers. 
                 The detection efficiency is given as a red curve. 
                 Larger occulter sizes, $r$, produce events that are easier to detect. 
                 Even though events with $r\approx 2$\,FSU are detected more than 50\% of the cases, 
                 the smaller events down to $r\approx 0.5$\,FSU are detected less than 10\% of the time.  
                 Lower values of stellar size, $R_\star$, velocity, $v$, and impact parameter, $b$, 
                 result in higher detection rates. 
            }
        \label{fig: efficiency fractions}
        
    \end{figure*}

    The simulations also allow us to find what fraction of events 
    are recovered as a function of different input parameters. 
    We show the number of events and the number of triggered events, 
    for different parameters, in Figure~\ref{fig: efficiency fractions}. 
    The red line in each plot represents the fraction of events recovered. 
    As expected, events are preferentially detected 
    at high occulter radius $r$, 
    but at low stellar radius $R_\star$, velocity $v$ and impact parameter $b$. 
            
    To calculate the histograms in Figure~\ref{fig: efficiency fractions} we summed all events 
    that fit a particular bin for the relevant parameter, 
    with a mix of the other occultation parameters and stellar properties 
    (marginalizing over the other parameters). 
    To increase the detection rate and highlight only the effects 
    of changing one parameter at a time, we apply some pre-selection rules
    on the events that would be marginalized in each histogram. 
    The rules are $r\geq 1$\,FSU, $R_\star\leq 1$\,FSU, $b\leq 1$\,FSU, 
    and $v\leq 20$\,FSU\,s$^{-1}$. 
    For each histogram we skipped the rule for the free parameter and 
    used the remaining three criteria for selecting the events to plot. 
    For example, each bin in the occulter radius distribution includes
    events generated with randomly chosen impact parameters with $b\leq 1$\,FSU, 
    randomly chosen velocities with $v\leq 20$\,FSU\,s$^{-1}$, 
    and stellar radii and photometric $S/N$ values of real star observations
    in our dataset, but with $R_\star\leq 1$\,FSU. 
    For the remaining histograms we limited $r\geq 1$\,FSU. 
    The occulter radius for simulated events was chosen randomly 
    from a power-law distribution: $dN/dr\propto r^{-3.5}$, 
    which is not necessarily the best estimate for the true 
    KBO distribution, but provides a good approximation
    that generates more shallow events and few deep events, 
    making sure we get sufficient events for a good statistical sample 
    of triggered events in all size bins. 
    The specific distribution of occulter sizes 
    has no effect when calculating the 
    detection efficiency as a function of occulter radius, 
    since in that case we calculate the efficiency for each 
    size bin separately. 
    
    The occulter radius distribution shows a steady increase 
    of detection efficiency with size, 
    from single-digit percentages at $r<0.5$\,FSU, 
    through 10--20\% detection rate at moderate sizes $r\approx 0.5-1$\,FSU
    and growing above 50\% for $\approx 2$\,FSU. 
    
\begin{figure}
        
        \centering
        
        \pic[1]{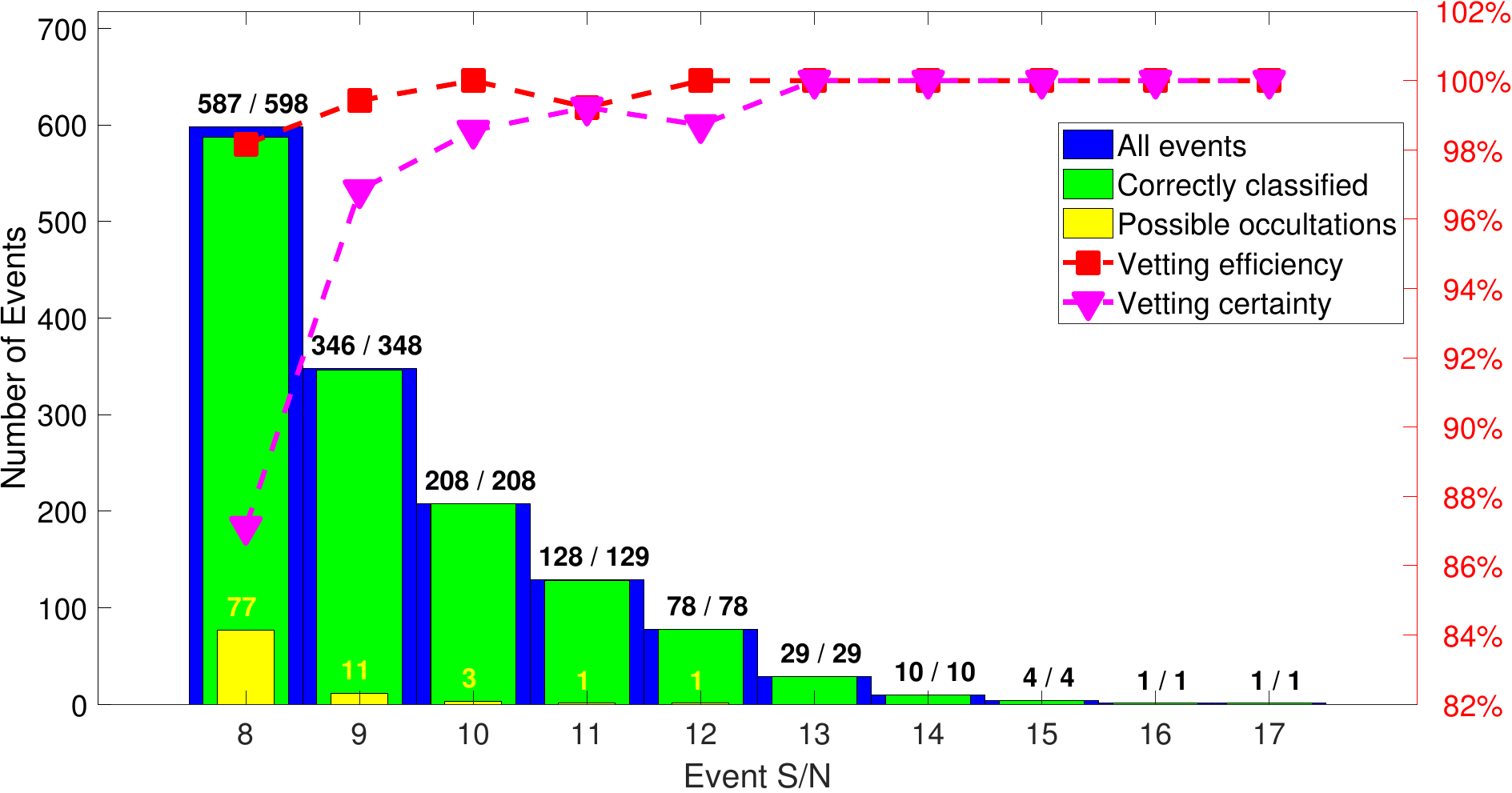}
        \caption{The number of simulated events that triggered and were not rejected
                 by data quality cuts, as a function of event $S/N$, 
                 are shown as blue bars.                  
                 The green bars show the number of events correctly classified. 
                 The red squares represent the fraction of correct classifications. 
                 Even for low $S/N$ the detection rate is close to 98\%. 
                 The yellow bars show the number of events classified
                 as ``possible occultations'', 
                 and the magenta triangles show the fraction 
                 of such classifications out of all triggered, 
                 simulated events 
                 (using the same right vertical axis as the red squares).
                 In this case we see that for low values of $S/N<8.5$ 
                 around 12\% of the events are classified as
                 possible (rather than certain) occultations. 
        }
        \label{fig: vetting fractions}
        
        
    \end{figure}

    In Figure~\ref{fig: vetting fractions} we show the 
    number of detected simulated events, 
    and also how many of them were correctly classified as occultations, 
    as a function of the event $S/N$. 
    The simulated events were injected into data
    collected from 2021 April to 2021 October,
    including $\approx 3\times 10^7$ individual exposures
    taken over 100 nights.     
    The squares show the fraction of events
    that were correctly classified. 
    We see that even for low $S/N$ events, 
    the losses due to human vetting, 
    at least for simulated events, 
    are lower than 2\%. 
    Also shown, in yellow bars,  are the number of ``possible occultations'', 
    those events for which the human vetter was not sure
    if the event is truly an occultation. 
    For non-simulated events that would be tagged as a possible occultation, 
    additional checks should be conducted to verify if the event
    is real or just a noise artefact (see \citealt{WFAST_KBO_search_limits_Nir_2023}). 
    The fraction of those events 
    is shown as magenta triangles. 
    The number of events classified as possible occultations
    is around 12\% for low $S/N$ ($<8.5$). 
    
    \subsection{Binary stars}\label{sec: binary stars}
    
    In this analysis we did not treat the case 
    where the occulted star	is in an unresolved binary. 
    The angular Fresnel scale for KBOs is $\sim 40$\,$\mu$arcsec, 
    which is smaller than the separation of 0.04\,AU at 1\,kpc. 
    Since $<1$\% of binaries have such a small separation 
    \citep{stellar_multiplicity_Duquennoy_Mayor_1991, stellar_multiplicity_Raghavan_2010}, 
    in most cases we expect the occulter to pass 
    in front of one of the two stars, leaving the other star's flux unaffected. 
    If the stars have similar fluxes, the occultation depth will be shallower relative 
    to the simulated depth for a single star, 
    and the occultation could display broadening due to the larger
    effective stellar radius $R_\star$. 
        
    While binary stars are common, 
    their masses are often unequal. 
    The mass ratio $q_m$ is constant for low $q_m<0.3$ 
    and drops faster at higher values of $q_m$
    \citep{binary_mass_ratio_Hogeveen_1992,binary_mass_ratio_Ducati_2011,binary_mass_ratio_Maxwell_2017}. 
    Since the broadband luminosity of a star is proportional to 
    a high power of its mass $L\propto m^{3.5}$, 
    we expect the vast majority of binaries to have
    very different flux values. 
    For a uniform mass ratio only 3\% 
    of binaries would have a secondary that contributes
    more than 10\% of the light. 
    If the bright star is occulted, 
    the difference in occultation depth would be small in most cases. 
    If the faint star is occulted, 
    the occultation would not be detectable. 
    
    It should be noted, however, 
    that \cite{binary_mass_ratio_Maxwell_2017}
    estimated a non-negligible fraction of binaries 
    with nearly equal mass, 
    mostly for solar-type stars. 
    While this population could reduce our
    detection efficiency by a few percent, 
    an exact estimate of this effect is 
    beyond the scope of this work. 
    
    In any case, once occultations are detected 
    the occulted stars can be studied further, 
    correcting the occultation parameters based on
    the specific properties of the system if 
    it is composed of multiple stars.

\section{Simulating occultation light-curves}\label{sec: simulations}
    
    Simulated light-curves are used extensively by this analysis pipeline:
    they are required for producing the template banks for the matched-filtering of light-curves (see \S\ref{sec: matched filtering}),
    they are injected into the data to estimate our detection efficiency (see \S\ref{sec: injection simulations}),
    and they are used to fit the parameters of each detected event (see \S\ref{sec: parameter estimation}).
    
    Thus, it was necessary to develop an efficient simulator that can produce 
    occultation events quickly and across a range of possible parameters. 
    A list of the different parameters used in our simulations is given in 
    Table~\ref{tab: occultation parameters}. 
    In this Section we describe the simulator algorithm, 
    its advantages for quickly producing light-curves,
    and its limitations. 
    
    	\begin{table}
    		
    		\centering
      
    		\caption{Short-hands used to describe observational and occultation parameters. 
    			Most parameters are given in units of the Fresnel Scale Unit (FSU).
    			The range of each parameter is given, 
                    denoting the possible values allowed by the simulation.
    			Where a single value is given, all simulations are run with just one parameter value. }
    		\begin{tabular}{l|c|c}
    			Parameter & Shorthand & Units \\ \hline \hline
    			
    			Stellar radius                        & $0<R_\star<10$       & FSU \\ 
    			Occulter radius                       & $0.1<r<3$            & FSU \\
    			Impact parameter                      & $0<b<2$              & FSU \\ 
    			Transverse velocity                   & $3<v<30$             & FSU/second \\
    			Mid-occultation time                  & $t_0=0$              & seconds \\ \hline
    			Integration time                      & $T=0.04$ or $T=0.1$  & seconds \\
    			Frame rate                            & $f=25$ or $f=10$     & Hz \\ 
    			Time window                           & $W=8$                & seconds 
    			
    		\end{tabular}
    		
    		\label{tab: occultation parameters}	
    			
    	\end{table}

    \subsection{Theoretical occultation light-curves}\label{sec: theoretical lightcurves}
    
\begin{figure}
        
        \centering
        
        \pic[1]{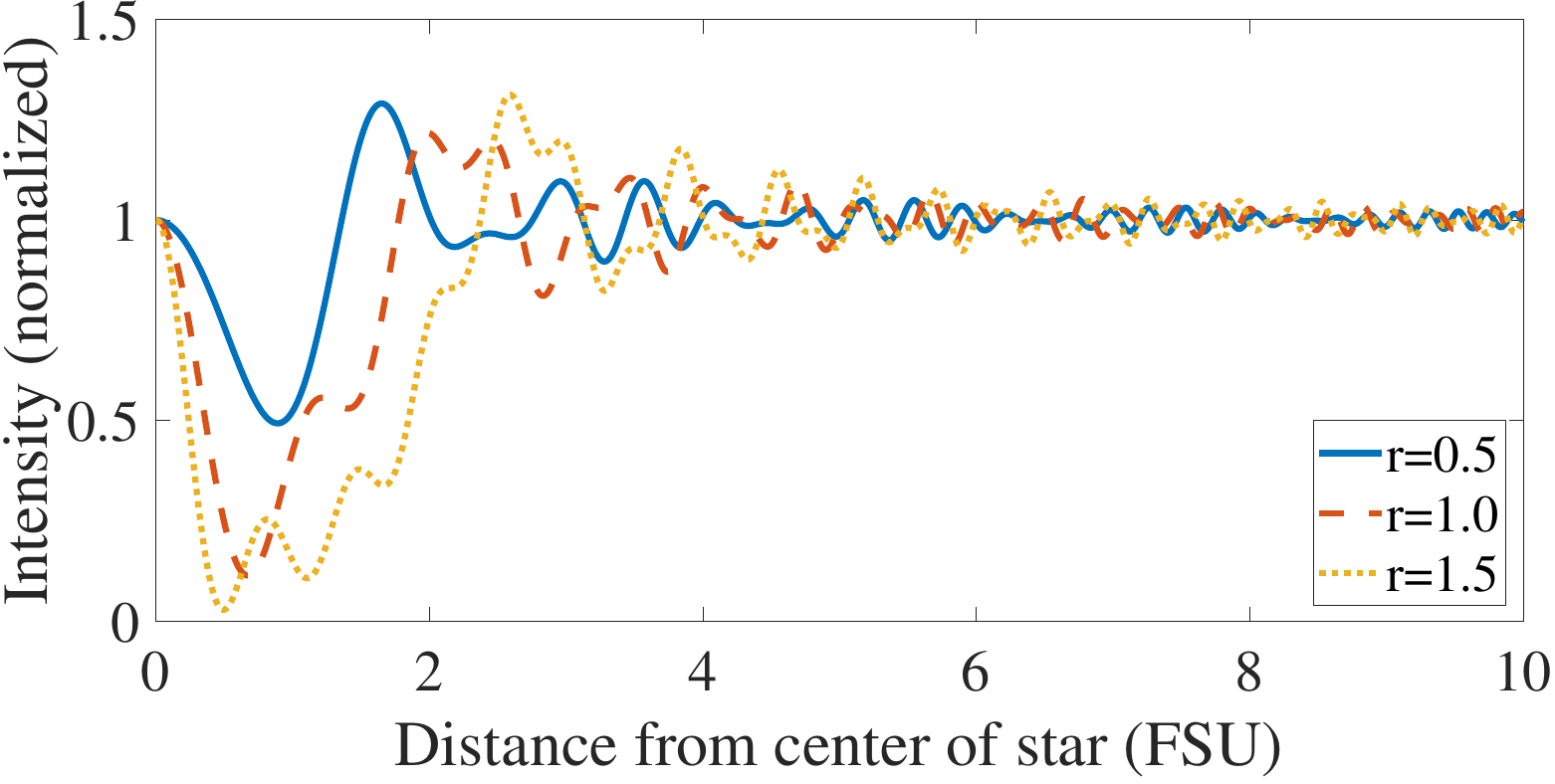}
        \caption{Examples of high-resolution illumination patterns for a few values of $r$, with $b=0$ and $R_\star=0$. 
                 The occulter radius is in Fresnel scale units. }
        \label{fig: example source matrix}
        
        
    \end{figure}

    The amplitude of light is calculated using \cite{KBO_occultation_diffraction_Roques_1987}, Eq.~9, 
    given here in natural units (using the Fresnel scale):
    \begin{equation}\label{eq: roques integral}
        a = 1 + i\pi e^{\frac{i\pi \rho^2}{2}} \int_0^{r} e^{\frac{i\pi r'^2}{2}} J_0(\pi r'\rho) r' dr',
    \end{equation}
    where $\rho$ is the distance between the center of the illumination pattern and an arbitrary point 
    (the distribution is circularly symmetric if the occulter is spherical),
    while $J_0$ is the zero order Bessel function $J_0(x)=\frac{1}{\pi}\int_0^\pi \cos(x \sin t) dt$. 
    These integrals were calculated once, for every occulter radius in the range $0.1<r<3$ in steps of 0.1 
    and for distances $0<\rho<35$ in steps of 0.0025. 
    The light intensity is then calculated using $I=|a|^2$ for each point in the illumination pattern. 
    These light-curves represent a high-resolution sampling of the intensity 
    at different $\rho$ values along the radial direction, measured from the center of the illumination pattern. 
    Such a light-curve is calculated for each $r$ value, and later used as the basis for all further calculations. 
    A few example light-curves for different occulter radii are shown in Figure~\ref{fig: example source matrix}.
    
    \subsection{Finite star sizes}\label{sec: star size}
    
\begin{figure}
        
        \centering
        
        \pic[1]{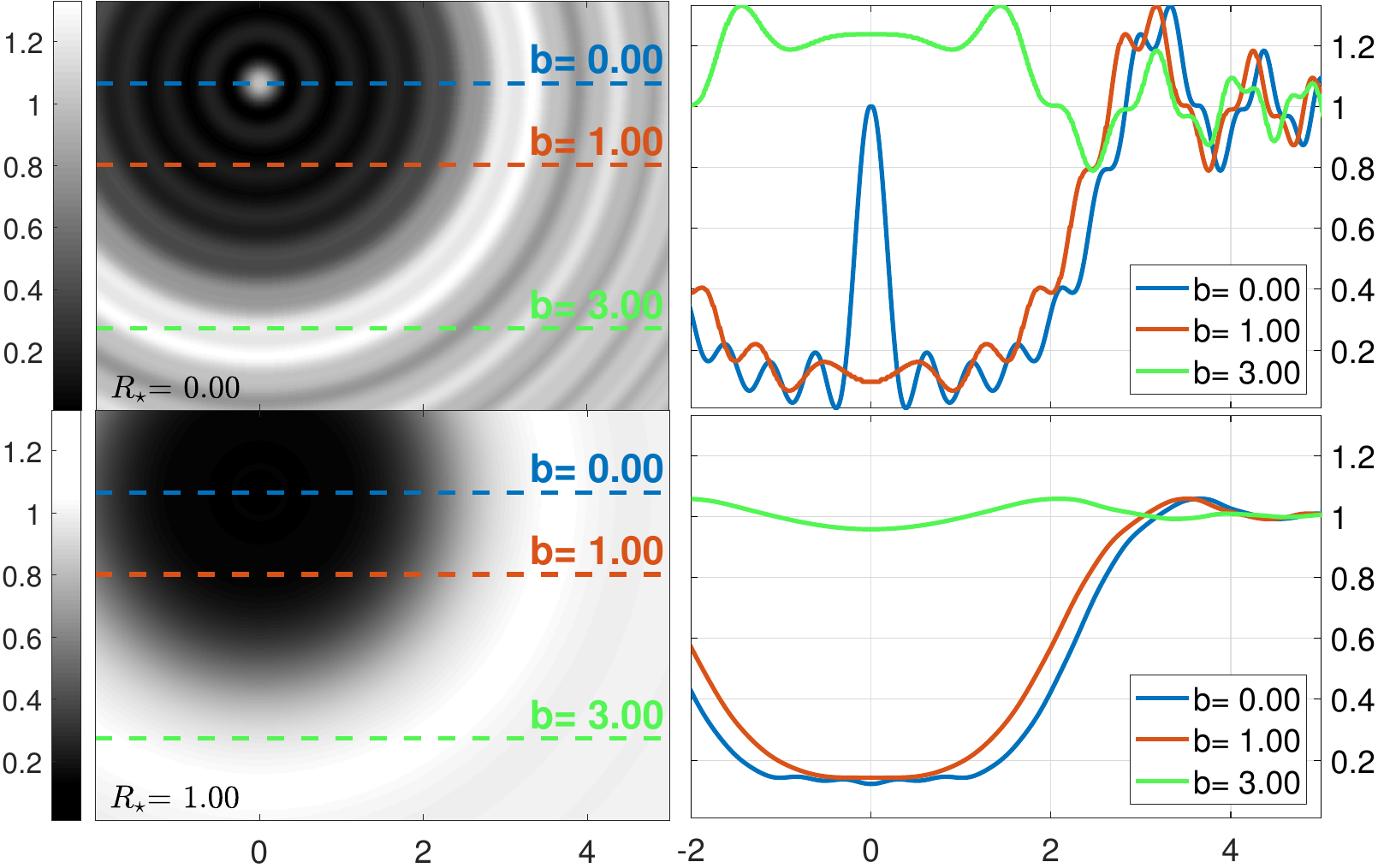}
        \caption{Examples for the 2D illumination pattern maps for the cases of 
            $R_\star=0$ (top) and $R_\star=1.0$\,FSU (bottom). 
            The left panels show the illumination patterns, 
            where the dashed, colored lines represent events with 
            different impact parameters. 
            The right panels show the resulting light-curves for such events. 
            The non-zero stellar radius smears out the illumination pattern, 
            weakening the strongest features in the light-curve and obliterating the weaker ones, 
            e.g., the Poisson peak in the $R_\star=0, b=0$ curve. 
            The curve with $R_\star=0, b=3.0$\,FSU 
            passes in a region of higher flux, 
            leading to a flare instead of a dip in the light-curve. 
            The spatial axes are in FSU, 
            and the color scale, 
            as well as the right-hand plot $y$-axis
            are given in relative flux units
            with the same intensity range. 
            }
        \label{fig: example illumination maps}
        
        
    \end{figure}

    A background star with finite (non-zero) radius creates an illumination pattern 
    which is the incoherent sum of the illumination patterns of each point on its surface. 
    Thus, the effects of non-point-source stars can be simulated by convolving the 
    two-dimensional illumination pattern for $R_\star=0$ with an image of the star, 
    a uniform circle in this case.\footnote{In future work we may implement limb-darkening as well.}
    The one-dimensional light-curves that were calculated for a point-source are imaged on a 2D map, 
    then 2D convolution smears the pattern by the correct amount using a circle of radius $R_\star$. 
    The star radius is calculated in Fresnel-scale units, 
    projected to the distance of the occulter (for KBOs, we use $D=40$\,AU). 
    This is equivalent to comparing the angular size of the star to the angular size of the occulter. 
    Example 2D maps of the illumination pattern with and without smearing 
    by an $R_\star=0.3$\,FSU star are shown in Figure~\ref{fig: example illumination maps}. 
    The results for different values in the range $0<R_\star<10$\,FSU in steps of 0.25 are also saved for 
    each occulter radius $r$ and distance $\rho$, 
    making a 3D matrix that is the source for all subsequent calculations. 
    We refer to this matrix as the \textit{source matrix}. 
    This matrix is generated once, taking a few hours, and then saved to disk. 
    From this matrix, high-resolution light-curves can be drawn for arbitrary values
    of $R_\star$ and $r$ (within the ranges given) by a linear interpolation between nearby grid points. 
    
    \subsection{Time and distance remapping}\label{sec: time remapping}
    
    The impact parameter $b$ has two effects on the final light-curve:
    (a) it skips the central region of the illumination pattern (i.e., the Poisson peak); 
    (b) it increases the time it takes to move from one diffraction crest/null to another when close to the point of nearest approach. 
    The sampling points at distances $a$ on the 1D source light-curve are given by $a=\sqrt{b^2+(vt)^2}$, 
    where $v$ is the velocity and $t$ is the time, measured from the mid-point of the occultation. 
    The light-curve is mirrored around the point of closest approach to make a symmetrical, 
    high-resolution light-curve of the entire event. 
    
    The remapped light-curve is integrated according to 
    the observation frame-rate, $f$, and integration time, $T$,
    with the appropriate time offset, $t_0$ (usually taken as zero for simplicity)\footnote{
        In the simulations conducted for this work, we used $t=0$ for all lightcurves. 
        In future work, we intend to try allowing $t$ to vary as a free parameter. }. 
    The observation window, $W=8$\,s, 
    determines the length of time for which the light-curve is calculated. 
    
\begin{figure}
        
        \centering
        
        \pic[1]{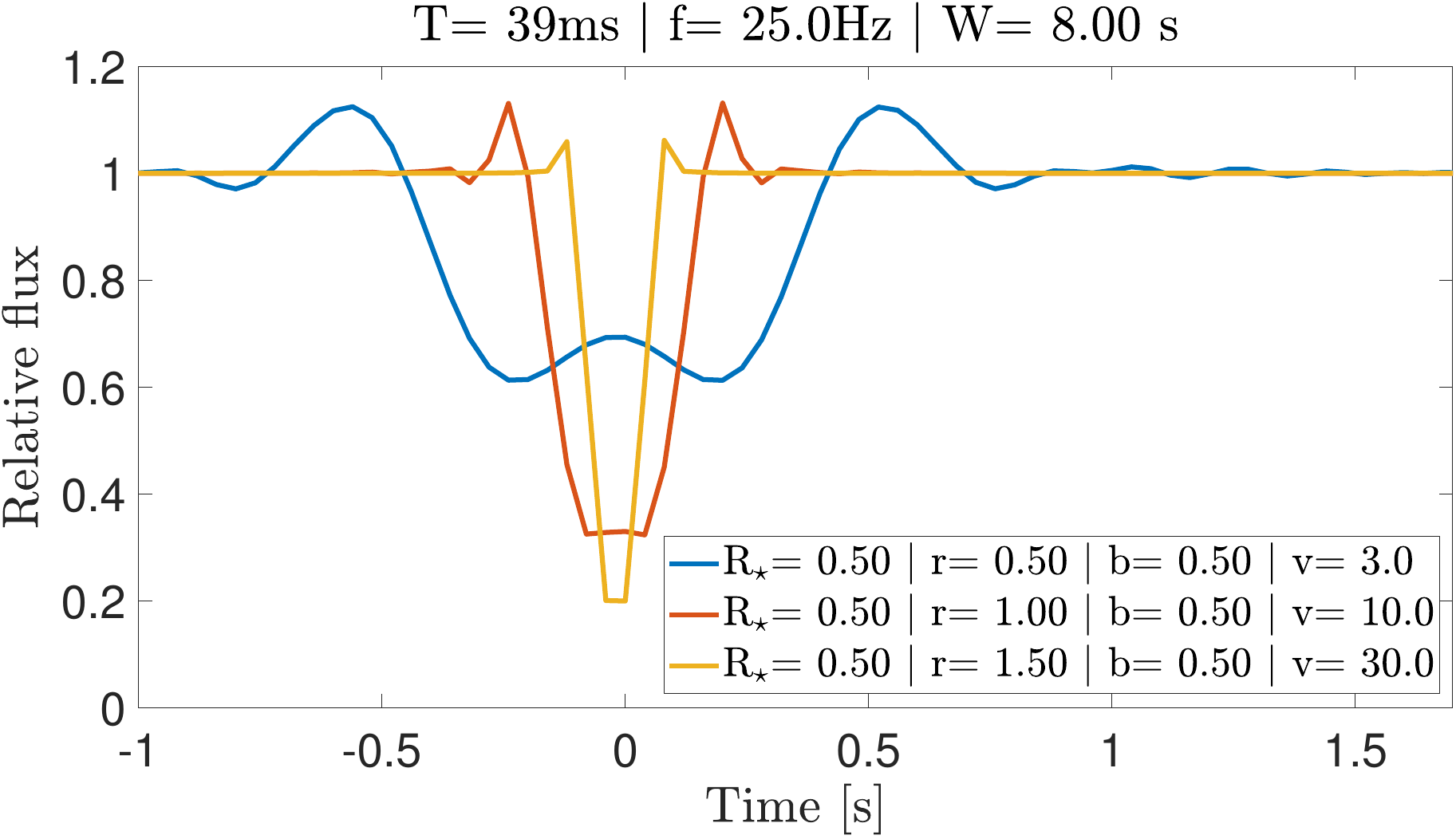}
        \caption{Examples of the final simulated light-curves for a few different parameter combinations.
            Events with $r\geq 1$\,FSU are seen to be deeper than the shallow $r=0.5$\,FSU case. 
            On the other hand, the low velocity of the $r=0.5$\,FSU event makes it much wider. }
        \label{fig: example lightcurves}
        
        
    \end{figure}

    For all simulations, the values $f=25$\,Hz and $T=40$\,ms are chosen to mirror the nominal values used by W-FAST. 
    For occultations with high $v$, the time window is larger than the entire re-sampled light-curve. 
    In these cases, unity flux values are added to the edges of the light-curve to represent the constant, un-occulted light-curve of the star. 
    A few examples for the final light-curves, after time-remapping and binning, are shown in Figure~\ref{fig: example lightcurves}. 
    
    \subsection{Bandwidth of visible light}\label{sec: broadband light}
    
\begin{figure}
        
        \centering
        
        \pic[1]{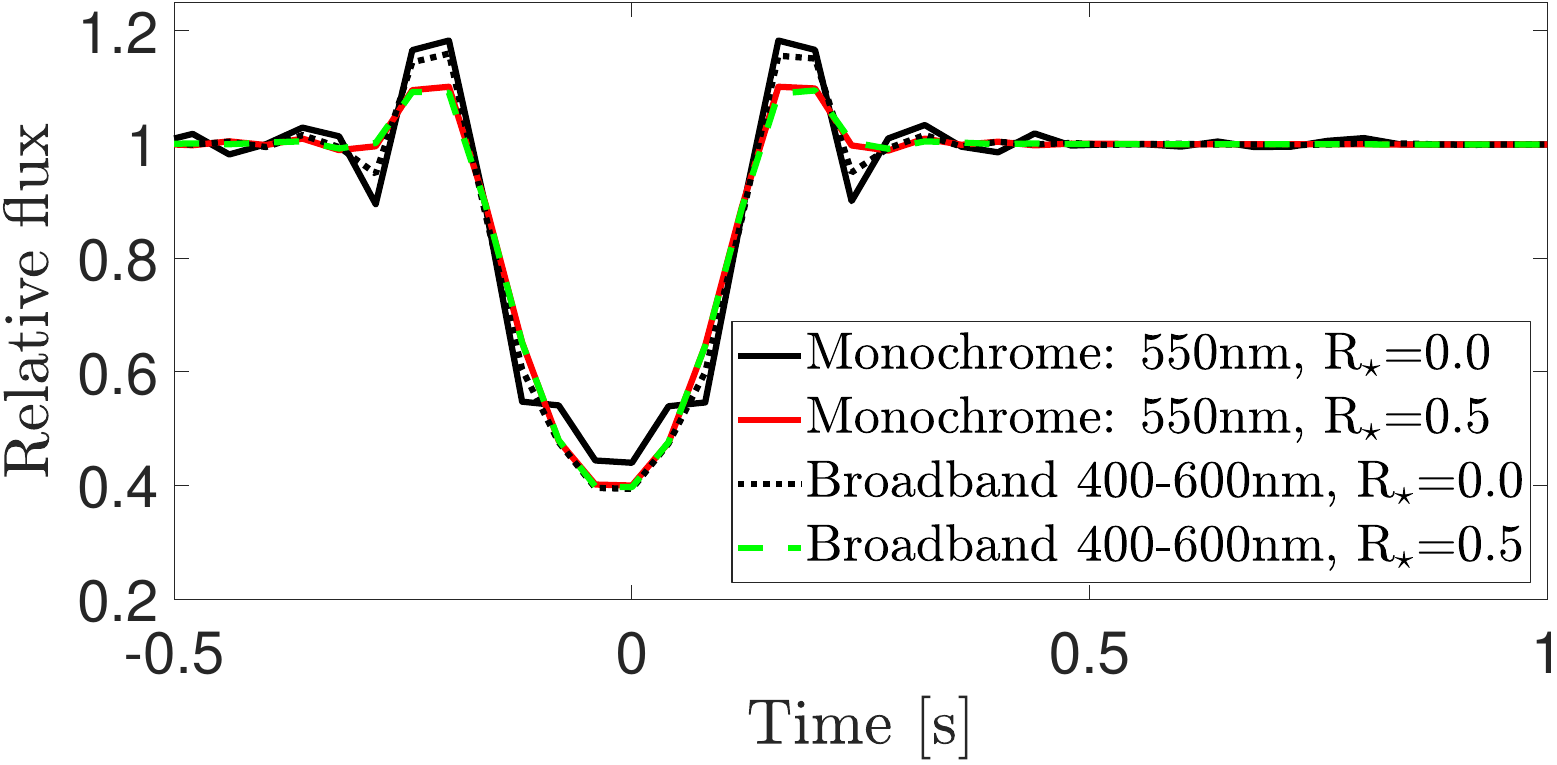}
        \caption{Light-curves for a $r=1$\,FSU, 
            $b=1$\,FSU and $v=10$\,FSU\,s$^{-1}$ occultation. 
            The different plots show the resulting light-curve for a point-source star and an $R_\star=0.5$\,FSU star, 
            and for monochromatic light (at 550\,nm) and for broadband light (400-600\,nm). 
            The light-curves for point-source stars are somewhat similar for monochromatic and broadband light
            (solid, black line vs.~dashed, black line), with the broadband being more smoothed in the central region. 
            The light-curves for the finite-size star (solid, red curve and dashed, green curve) are both 
            more smoothed in the central region but also on the diffraction fringes. 
            For $R_\star=0.5$\,FSU the effects of using broadband light is negligible. }
        \label{fig: broadband lightcurves}
        
        
    \end{figure}

    The light-curves thus far are calculated for monochromatic light. 
    The effects of non monochromatic light on the illumination pattern is similar 
    to that of non-point source stars.
    The light from various points on the star do not interfere, 
    and neither does light at different wavelengths: 
    in both cases multiple illumination patterns are shifted and superimposed on each other incoherently. 	
    Since both effects cause the light-curves to be smeared out, 
    without dramatically changing the detection statistics, 
    treating them together does not make a big difference. 
    As an example, light-curves for monochromatic and broadband light
    are shown in Figure~\ref{fig: broadband lightcurves}. 
    
    The finite bandwidth effect is small compared 
    to the smearing caused by the finite size of stars. 
    We neglect this effect in our simulations, 
    assuming it is included in the stellar size smearing
    and that any remaining effects are negligible. 
    
    \subsection{Geometric approximation for large stars}\label{sec: geometric approximation}
    
\begin{figure}
    
        \centering
        
        \pic[1]{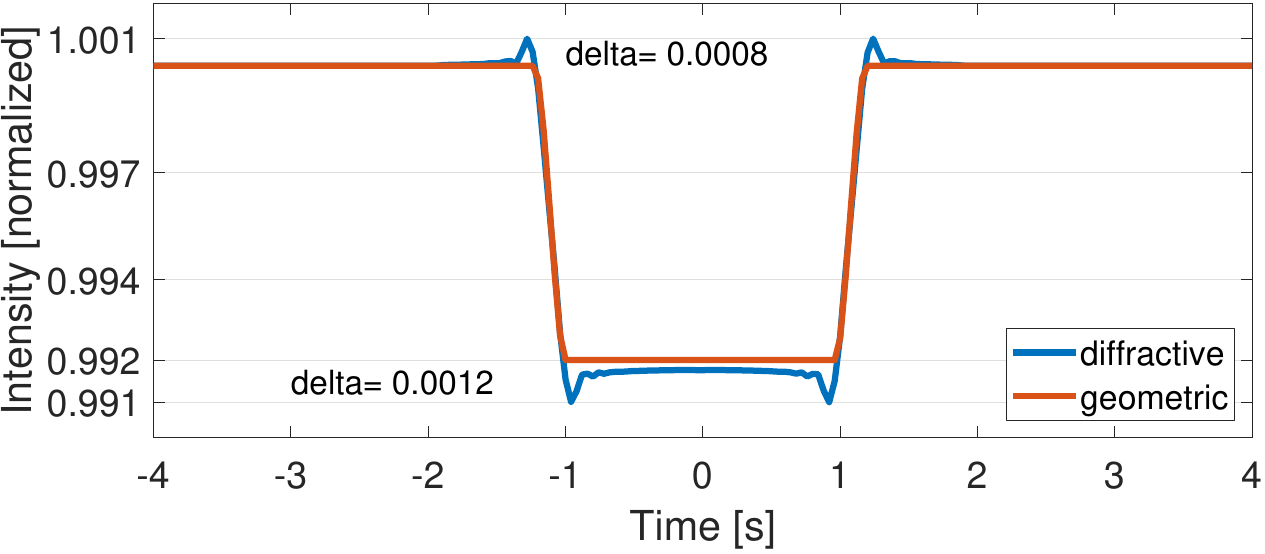}
        \caption{Two light-curves with identical parameters, 
            produced using a diffractive and a geometric simulation. 
            The parameters are $R_\star=11$, $r=1$, $b=1$, and $v=10$, 
            all in Fresnel units. 
            The differences appear at the entrance and exit into partial and full shadow. 
            The sizes of the differences, relative to the light of the star, 
            are on the order of $10^{-3}$, well below the noise level 
            in our data. 
        }
        \label{fig: geometric comparison}
        
    
    \end{figure}

    For computational reasons, 
    for stars with $R_\star>10$\,FSU we use a geometric approximation 
    instead of the diffractive methods described in previous sections.
    For such large stars the diffractive features are washed out, 
    and there are only minor differences between geometric and diffractive occultations. 
    It is evident that as the stellar size grows, the light-curve 
    transitions fairly smoothly into the geometric regime. 
    Under normal noise conditions, the differences are negligible. 
    In Figure~\ref{fig: geometric comparison} we show
    two light-curves with the same parameters, 
    one produced using the full diffractive simulation, 
    and the other using the geometric approximation. 
    The maximum difference between the light-curves is $\approx 1.2\times 10^{-3}$.
    
    We test the difference between the geometric and diffractive simulation methods  
    as a function of different parameters, 
    recording for each pair of light-curves 
    the maximal flux difference across the entire light-curve. 
    We chose nominal values of $R_\star=10$\,FSU, 
    $v=10$\,FSU\,s$^{-1}$ and varied the values of $r$ and $b$. 
    We used a constant $b=1$\,FSU when varying $r$, 
    and a constant $r=2$\,FSU when varying $b$. 
    We find that the differences never surpass 1\%, 
    which is well below the noise floor for all of our measurements.

    \subsection{Binary occulters}\label{sec: binary occulters}
    
    Observations of larger KBOs ($r>10$\,km) have shown that many
    objects are found in binaries \citep{kuiper_belt_binaries_Kern_Elliot_2006, kuiper_belt_binaries_Schlichting_Sari_2008, KBO_binaries_Showalter_2021}, 
    or have a bi-modal shape (e.g., Arrokoth; see also \citealt{kuiper_belt_binaries_Gnat_and_Sari_2010}). 
    The bi-modal case is equivalent to a binary occulter observed
    at a time when the objects are partially obstructing each other 
    and their projection forms a continuous shape. 
    In either case the occulter is no longer circular, 
    and specific methods should be used to simulate the light-curves from such a configuration
    (e.g., \citealt{occultation_simulator_TAOS_Castro_2019}). 
    
\begin{figure}
        
        \centering
        
        \pic[1]{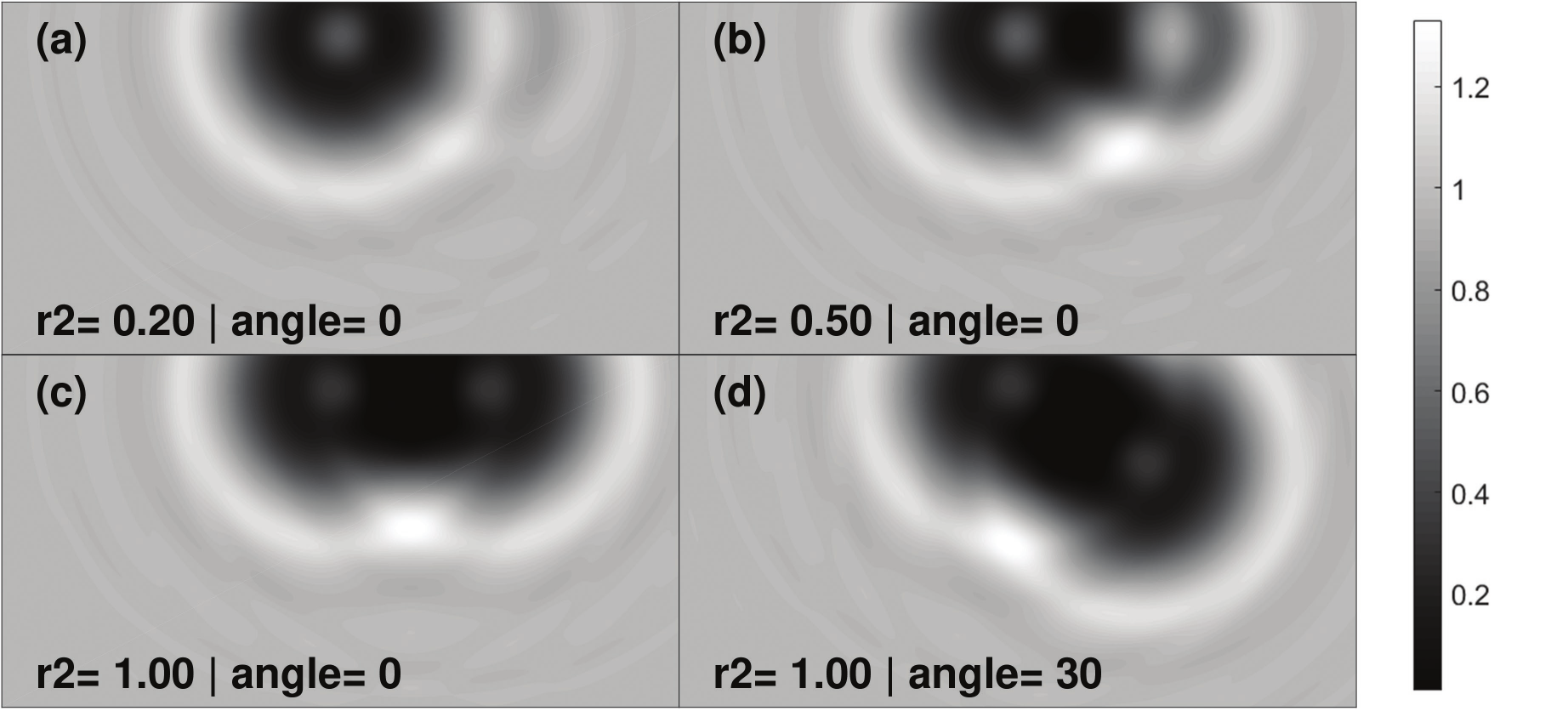}
        \caption{Intensity maps for occultations by binary objects. 
            The color map scales from 0.2 (black) to 1.3 (white) of the star's flux. 
            (a) the secondary with $r_2=0.2$\,FSU is barely noticeable; 
            (b) the secondary with $r_2=0.5$\,FSU has a strong effect on the map; 
            (c) the secondary has the same size as the primary $r=r_2=1$\,FSU, 
            generating a symmetrical intensity map;
            (d) the same occulter sizes are seen at an angle of $\theta=30$ degrees, 
            breaking the symmetry for an observer seeing the event unfold along the $x$ axis. 
        }
        \label{fig: binary intensities}
        
        
    \end{figure}

    When simulating events with two occulters,
    a secondary simulation method was used. 
    In this method the light field amplitude was calculated using Equation~\ref{eq: roques integral}, 
    and translated to a 2D map around the center of the occultation. 
    A companion can be added by generating a second 2D amplitude map, 
    with its center shifted by some distance $d$ in FSU, and some angle $\theta$ in degrees. 
    The complex amplitudes are added coherently, 
    then taken in absolute square to yield the intensity map. 		
    If a non zero $R_\star$ is used, the intensity map is convolved with a circle of the correct size, 
    as explained in \S\ref{sec: star size}. 
    Some example intensity maps for binary occulters are shown in Figure~\ref{fig: binary intensities}. 
    
    This method is slower than using the source matrix but is useful
    for simulating binary or bi-modal occulters. 
    A limitation of this method is that the two occulters must not overlap:
    the integral given in Equation~\ref{eq: roques integral}
    assumes a circular shape, and any overlap between objects is not treated correctly in this way.
    To simulate contact binaries, bi-modal objects and partial eclipses in detached binaries, 
    other methods must be used (e.g., \citealt{occultation_simulator_TAOS_Castro_2019}). 
    
    To get a sense of the size of the effect that the existence 
    of a secondary occulter would have on the light-curves
    we simulated a few test cases. 
    In these simulations the main occulter had $r=1$\,FSU, 
    the impact parameter was $b=0$ and the velocity was $v=5$\,FSU\,s$^{-1}$ 
    to make the light-curves well sampled and easier to interpret. 
    The stellar radius was set to $R_\star=0.5\,FSU$. 
    The distance between occulters was taken as $d=2$\,FSU so that in all cases there 
    was no overlap between occulters. 
    We tested several companion sizes $r_2=0.2, 0.5,$ and 1\,FSU, 
    and also two angles, $\theta=0$ and 30\,deg between the line connecting 
    the occulters and the direction of their movement on the sky. 
    The resulting amplitude maps are shown in Figure~\ref{fig: binary intensities}. 
    In the case of $r_2=0.2$\,FSU it is hard to see the effect of the secondary at all. 
    In the other cases the additional occulter dramatically affects the intensity map. 
    
\begin{figure}
        
        \centering
        
        \pic[1]{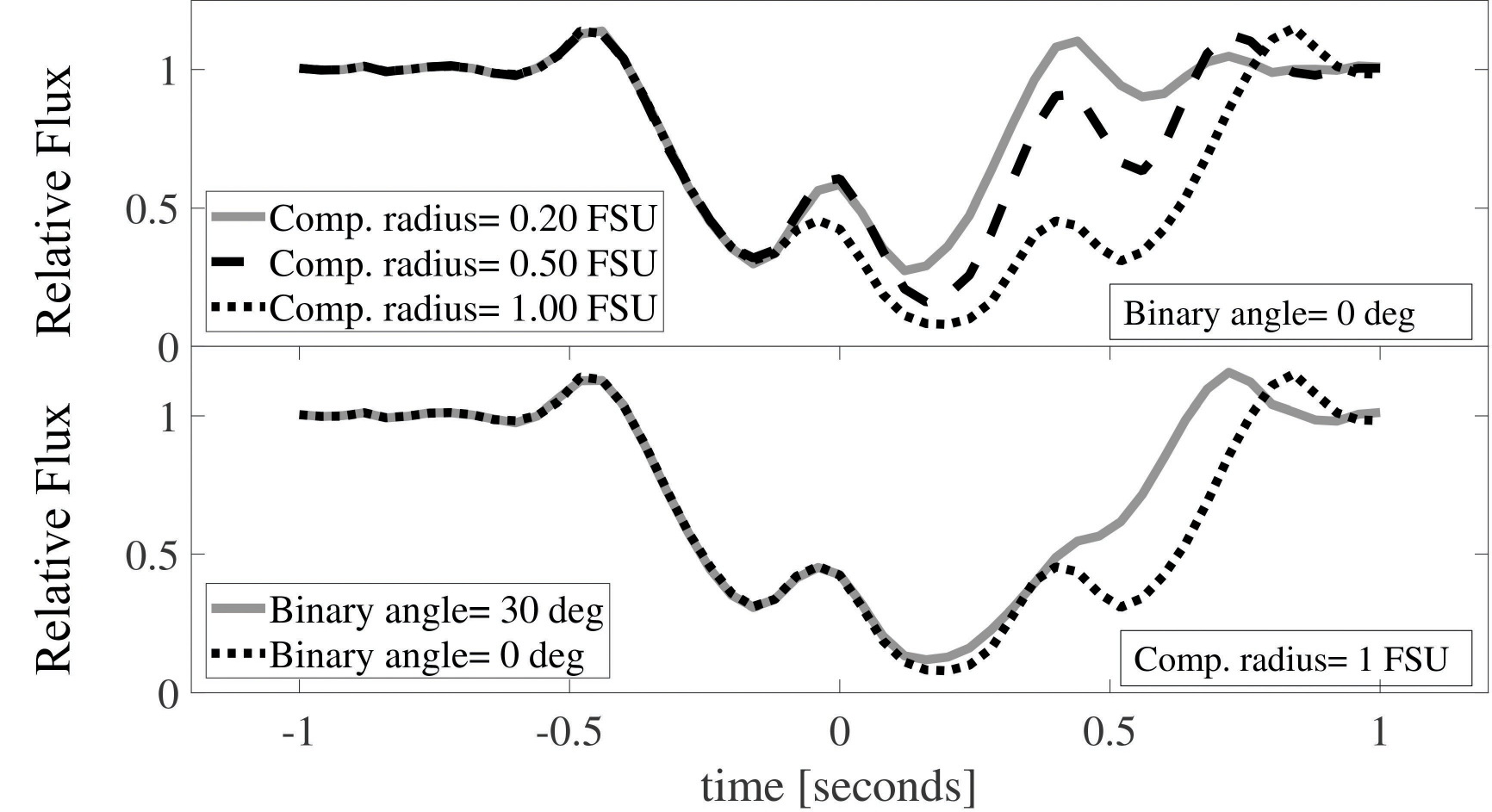}
        \caption{The light-curves for a few instances of binary occulters.
            All events have $R_\star=0.5\,FSU$, $v=5$\,FSU\,s$^{-1}$ 
            and a binary separation of $d=2$\,FSU. 
            In the top panel several occultations are shown at an angle $\theta=0$, 
            so the line connecting the centers of the occulters is parallel to their velocity. 
            The different secondary sizes $r_2=0.2, 0.5,$ and 1\,FSU show different effects. 
            The smallest secondary is practically invisible, 
            the intermediate secondary affects the light-curve in a noticeable and asymmetric way, 
            while the largest secondary is identical to the primary, inducing a symmetric, longer occultation. 
            On the bottom panel the two events with $r=r_2$ are shown for different angles
            between the line connecting the centers of the occulters and the velocity. 
            The dotted line shows the same symmetric light-curve from the top panel, 
            while the solid, gray line shows the asymmetric light-curve induced by the non zero angle. 
        }
        \label{fig: binary lightcurves}
        
        
    \end{figure}

    The light-curves produced by these events are shown in Figure~\ref{fig: binary lightcurves}. 
    The top panel shows several companion radii for the same angle $\theta=0$\,deg. 
    The light-curve for $r_2=0.5$\,FSU shows clear asymmetry between the flux before and after 
    the center of the occultation. 
    The light-curve for $r_2=1$\,FSU is symmetric only because the two occulters are identical. 
    In the bottom panel we show that the case of $r_1=r_2$ only yields symmetric light-curves
    when the binary is moving at specific angles ($\theta=90$\,deg or, as in this case, $\theta=0$\,deg). 
    For other angles the light-curve is once again left-right asymmetric. 
    It is evident that the light-curve is symmetric 
    only when there is complete symmetry between the two occulters, 
    in size and in the position angle. 
    Since this is expected to be quite rare, 
    it is likely that any binary occulters (that have a non-negligible secondary radius)
    will be discernible from a single occulter based on light-curve asymmetry.  
    
\begin{figure}
        
        \centering
        
        \pic[1]{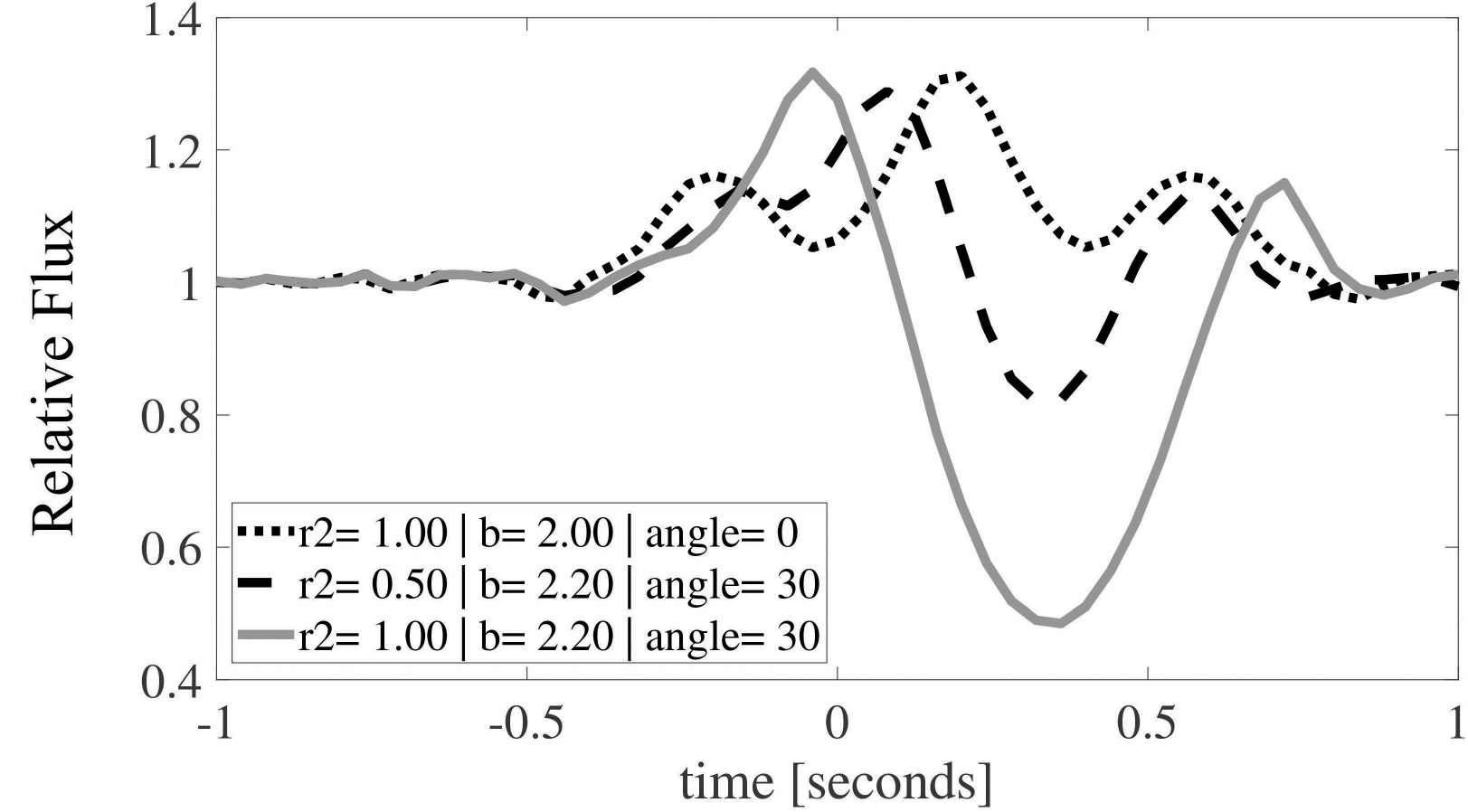}
        \caption{Simulated occultations for binary objects with a specific choice of impact parameter, $b$, 
            that highlights the potential for positive occultations to occur. 
            If the observer passes through an area in the intensity map where diffraction 
            causes a peak intensity, the star's light could be enhanced by a factor of 30\%. 
            In the solid, gray curve, and the dashed, black curve, 
            the occultations happen when the line connecting the centers of the occulters 
            is at an angle $\theta=30$\,deg to the velocity, so the light-curves also exhibit strong dips. 
            In these cases the secondary size affects the depth of the occultation dip. 
            The dotted line shows an occultation when the angle is $\theta=0$\,deg, 
            so that only the diffraction peak is visible, without any dimming of the star. 				     
        }
        \label{fig: binary flash}
        
        
    \end{figure}

    Furthermore, the effects of diffraction are evident in the intensity maps, 
    including bright spots that increase 
    the star's light by $\approx 30$\%. 
    A few example light-curves for such situations are shown in Figure~\ref{fig: binary flash}. 
    To generate these events, the impact parameter, $b$, 
    was chosen so the observation goes 
    through the diffraction peak in the intensity map. 
    In the cases when the line between the centers of the occulters 
    is at an angle $\theta=30$\,deg to the velocity trajectory, 
    the peak is followed by a large dip in the light-curve, 
    which depends on the size of the secondary. 
    In the case when $\theta=0$\,deg the observer traces a path
    that avoids any large dips and is exposed only to the diffraction peak. 
    In this case the light-curve shows a positive occultation without any dimming of the starlight. 
    In all the cases simulated, the peak brightness is $\approx 30$\% of the star's light. 
    This amplitude depends on the size of the occulters and the distance between them. 
    
    Finally, it is important to check the detection prospects of binary occultations. 
    We checked the compatibility of the light-curves presented in the above plots
    with the templates for single occultations. 
    Most cases of binary occultations presented in Figures~\ref{fig: binary lightcurves}
    and \ref{fig: binary flash}
    have an overlap (trigger efficiency) greater than 90\%, 
    indicating minimal loss of information and $S/N$ (see \S\ref{sec: template banks}), 
    even though the templates are based on single occulters. 
    The cases where the overlap is lower, but still above 80\%, 
    are when the occultation passes through the diffraction hot-spot. 
    Therefore, we conclude that even though there is a large parameter space
    of binary-occultation templates that can be generated, 
    triggering based on single-occulter templates does not cause large losses 
    to the trigger efficiency. 
    However, it is important to also trigger on brightening events, 
    not only on light-curve dips, as they could be caused
    by binary occultations, or other unknown physical phenomena. 
    
    If a binary occultation is detected, it would most likely 
    be easily identified as such because it is either 
    (a) asymmetric around the dip center or
    (b) displays brightening instead, or in addition, to dips in the light-curve. 
    If such events are detected, more binary simulations 
    could be performed to find the event parameters. 
    
    The case of binary occulters is only one possible extension 
    to the idealized case of a spherical occulter. 
    Other possible occultations may be caused by contact binaries 
    or binaries that are partially eclipsing at the time of the occultation, 
    as well as generally non-spherically shaped bodies. 
    The latter case can show similar shadow patterns to that of the close-
    or contact-binary case, if the object is bimodal or elliptical. 
    On the other hand, it may have a completely different, non-spherical shape, 
    e.g., a diamond shape similar to the asteroid Bennu. 
    If the occulter radius is small compared to the diffraction scale, 
    the specific shape should have a diffraction pattern 
    similar to the case of the spherical occulter. 
    If the object is larger than the diffraction scale, 
    we expect the influence of its shape to be more pronounced, 
    as in the case of the binary occulters. 
    In this case, however, it should also be easy to detect, 
    and would probably have a non-symmetric diffraction pattern, 
    so that its non-spherical nature would be apparent. 
    It may, however, be difficult to pin down the exact shape
    based solely on a single occultation. 
    
	\section{Parameter estimation}\label{sec: parameter estimation}
	
	After finding an occultation event, 
	we aim to constrain the underlying parameters, 
	namely the occulter size and orbital parameters. 
	To do so, we try to constrain the parameters 
	of the occultation itself, 
	including the stellar size $R_\star$, 
	the occulter size $r$, 
	impact parameter $b$
	and velocity $v$,
	from which we can derive
	the distance (assuming Keplerian motion). 
	We will assume, for simplicity, 
	that the occultation is perfectly centered at time $t=0$. 

    It should be noted that the occultation parameters do not directly 
    translate to the occulter's orbital parameters. 
    There is a degeneracy between the distance and velocity of the object
    and the occultation's depth and width. 
    For example, a main belt asteroid may present a similar
    occultation lightcurve as a KBO. 
    However, since KBOs are expected to have a higher angular number density 
    when searching $\pm 2$\,deg from the ecliptic 
    \citep{abundance_kuiper_objects_Schlichting_Ofek_2012}
    most of the objects that would be detected in 
    an occultation survey along the ecliptic plane would be KBOs.
    In future work, when robust detections of real occultations are made, 
    we plan to discuss ways to model the orbital parameters based on the occultation data. 

    For KBOs, we assume a semimajor axis of 40\,AU and zero eccentricity. 
    The real distribution of orbital parameters induces detection biases
    for objects with different orbital parameters. 
    However, the variation in the measured quantities between objects is not very large. 
    We estimate the velocity distribution of KBOs by simulating the momentary velocity 
    of all numbered KBOs (with semimajor axis above 35\,AU) at a hundred random dates, 
    and find $\approx 20$\% variation in the velocity, which does not greatly affect the detectability. 
    The variation in the distance only affects the Fresnel scale with a square root dependency, 
    so the variation of semimajor axes of KBOs will not dramatically change the results. 
 
	Using the simulation methods discussed in \S\ref{sec: simulations}
	we can produce simulated light-curves and compare them with the measured light-curve:
	\begin{equation}\label{eq: mcmc deltas}
		\chi^2 = \sum_{i=0}^N \left( \frac{f_{m,i}-f_{s,i}}{\sigma_i^2} \right)^2, 
	\end{equation}
	where $f_m$ and $f_s$ are the measured and simulated light-curves, respectively, 
	$\sigma_i$ is the RMS per measurement $i$, and the sum is over all $N$ samples in the light-curve. 
	For each simulated light-curve that we compare to the measured data
	we can calculate the logarithm of the likelihood for that data-set, given the parameters 
	used in the simulation:
	\begin{equation}\label{eq: chi square likelihood}
		LL = -\frac{1}{2} \chi^2.
	\end{equation}

	We use the Metropolis MCMC algorithm
	\citep{markov_chain_monte_carlo_Metropolis_1953}, 
	with random jumps in the parameter space using 
	a multivariate Gaussian with $\sigma$ parameters equal to
	0.1 for $R_\star$, 
	0.25 for $r$ and $b$, 
	and 1 for $v$ (all in Fresnel units).
	The allowed range of the parameters is 
	$0 \leq R_\star \leq 3 $\,FSU, 
	$0.1 \leq r \leq 3 $\,FSU, 
	$0 \leq b \leq 2$\,FSU, 
	and $1 \leq v \leq 40$\,FSU s$^{-1}$.
	If the walker tries to move out of the range of the parameters, 
	it ``bounces back'' elastically, 
	completing the randomly chosen step length along that parameter axis, 
	but in the opposite direction to how it was moving in the first part of the step.
	In each step we simulate a model occultation light-curve based
	on the randomly chosen parameters and compare it to the input flux
	using Equation~\ref{eq: mcmc deltas}. 
	The log-likelihood is calculated using Equation~\ref{eq: chi square likelihood}, 
	with an additional factor proportional to 
	\begin{equation}
	    LL_\text{prior} = -\frac{R_\star - R_{\star,\text{known}}}{2 (0.1 R_{\star,\text{known}})^2}, 
	\end{equation}
    where $R_{\star,\text{known}}$ is the best estimate of the target star's radius, 
    calculated by fitting a bolometric model to the star's Gaia colors, 
    converted to FSU. 
    This additional term represents a prior 
    (with typical width $0.1 R_{\star,\text{known}}$)
    that keeps the values of $R_\star$ of the chain points
    close to the best-estimate value.
    The other parameters do not directly affect the likelihood, 
    so the priors are uniform in the ranges given above. 
    
	We find the difference of log-likelihoods between the next point 
	and the previous point. 
	If the difference is positive (the likelihood ratio is larger than one)
	the new point is added to the chain. 
	If not, the difference is compared to the log of a random number, 
	uniformly distributed in the range $[0,1]$. 
	If the difference is larger than the randomly chosen number, 
	the new point is added to the chain. 
	If the difference is smaller, 
	the new point is rejected and the previous point is repeated in the chain. 
	We ignore a subset of points at the beginning of the chain, 
	which we call \emph{burn-in}, 
	to let each walker reach a part of the parameter space
	where the simulated light-curves show a good match to the data.
	All subsequent points after the burn-in are used to calculate the posteriors. 
	
    We run 10 different chains, 
    with starting positions chosen using a uniform draw
    of all four parameters in their respective ranges. 
	We run 1000 steps of burn-in, 
	and 9000 points that are used in the analysis. 
	For each chain we find the mean of $\chi^2$ values of all points, $\langle \chi^2 \rangle$. 
	The median of the average $\chi^2$ values between the different chains, $M_{\chi^2}$, 
	and the median absolute deviation of those values, $MAD_{\chi^2}$,
	are used to disqualify chains that
	got stuck in areas of low likelihood. 
	Any chain with 
    \begin{equation}
        \langle \chi^2 \rangle - M_{\chi^2} > 3 MAD_{\chi^2}
    \end{equation}
    is not used in finding the posterior distribution. 
	
\begin{figure}
		
		\centering
		
		\pic[1]{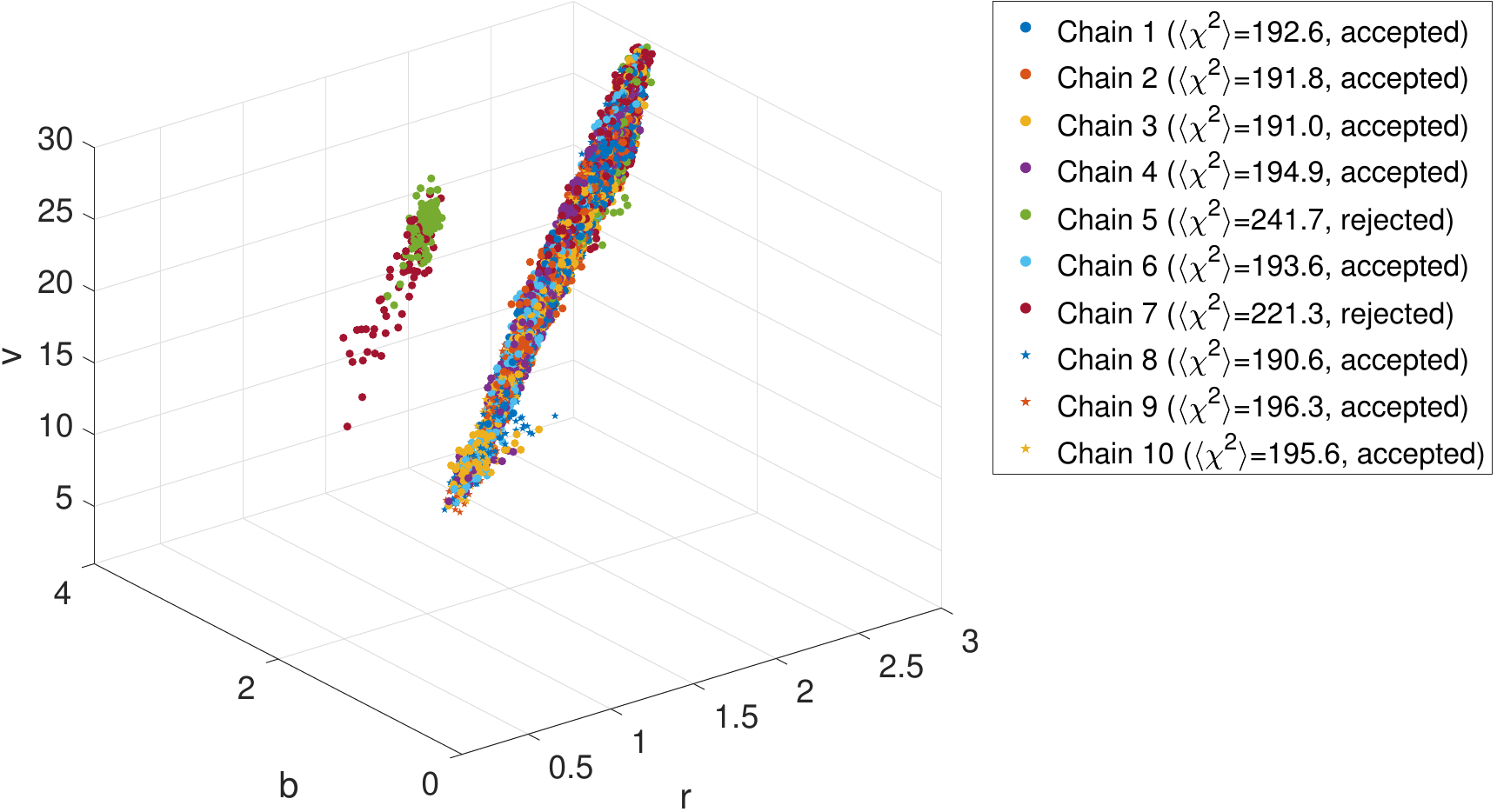}
		\caption{An example of 10 chains of MCMC for a single, simulated event. 
		         The light-curve and posteriors of this event are shown in Figure~\ref{fig: mcmc sim results 1}. 
		         The chains show a strong degeneracy between $r$, $b$ and $v$. 
		         Without narrow priors on $b$ and $v$, 
		         the range of possible $r$ values spans a large part of the parameter space, 
		         between 1.5 and 3\,FSU. 
		         There are two chains (numbers 5, and 7) that are 
		         offset from the other chains. 
		         These chains trace a local minimum with $\langle\chi^2\rangle\approx$ 220--240
		         which is less likely than the region traced by the other chains, 
		         with $\langle\chi^2\rangle\approx$ 190--200. 
		         Those two chains are rejected from the posterior analysis. 
		}
		\label{fig: mcmc example chains}
		
		
	\end{figure}

	The density of points of the remaining chains is used as an estimate
	for the posterior Probability Density Function (PDF). 
	Typically, there is a strong degeneracy between $r$, $b$ and $v$, 
	generating thin and long posteriors in $r$-$b$ plane, 
	or sheets in the $r$-$b$-$v$ space. 
	An example for such a distribution of points is shown 
	in Figure~\ref{fig: mcmc example chains}, 
    for a simulated event with 
    $R_\star=0.35$\,FSU, $r=1.7$\,FSU, $b=1.9$\,FSU and $v=25$\,FSU\,s$^{-1}$. 
	
	Since the impact parameter $b$ does not reflect on the physics
	of the occulter or its orbit, 
	we treat it as a nuisance parameter 
	and marginalize over it when finding the 
	distributions of the physical parameters, $r$ and $v$. 
	The star radius $R_\star$ is also not a physical property
	of the occulter and is generally well constrained 
	by the prior set by the Gaia data. 
	We show the posterior probability distributions for 
	a few sample, simulated events,
	in Figures~\ref{fig: mcmc sim results 1}--\ref{fig: mcmc sim results 3}.
	These events were injected into real data taken during 
	2021 while searching for real occultations. 
	We show the joint distribution of $r$ and $v$, 
	as well as the marginal probability distributions
	where we compare the best value and confidence limits 
	to the underlying simulation parameters. 
	
	The limits and best value are calculated for each 1D marginal posterior
	by finding the shortest parameter distance that encompasses 68\% of the points
	in the distribution. 
	Once the bounds are set, 
	we histogram the values inside the bounds into 100 bins. 
	The number of points in each bin is used as an initial estimate of the 
	probability distribution inside the bounds. 
	We smooth this distribution with a Gaussian
	with width of five bins, 
	and find the bin with the highest smoothed probability. 
	The center of this bin is taken as the best value. 
	Most of our distributions are fairly wide, 
	so the estimate for the best value is 
	not very precise. 
	However, we find that looking for the peak in the distribution (the mode)
	gives more reasonable results than the mean or median of the distribution, 
	which is often a-symmetric. 
		
	For real events, it is possible to compare the allowed range of velocities
	(e.g., the 68\% middle region of the distribution) 
	with the velocity physically allowed 
	for the observed field 
	based on the orbital motion of the Earth,
	projected onto the sky, 
	minus the expected velocity of a KBO on a circular or eccentric orbit at 40\,AU.  
	This provides an additional check for 
	distinguishing real occultations from spurious noise events. 

\newcommand{\mcmcfigsize}{0.72}
 
	\begin{figure*}
		
		\centering
		
		\pic[\mcmcfigsize]{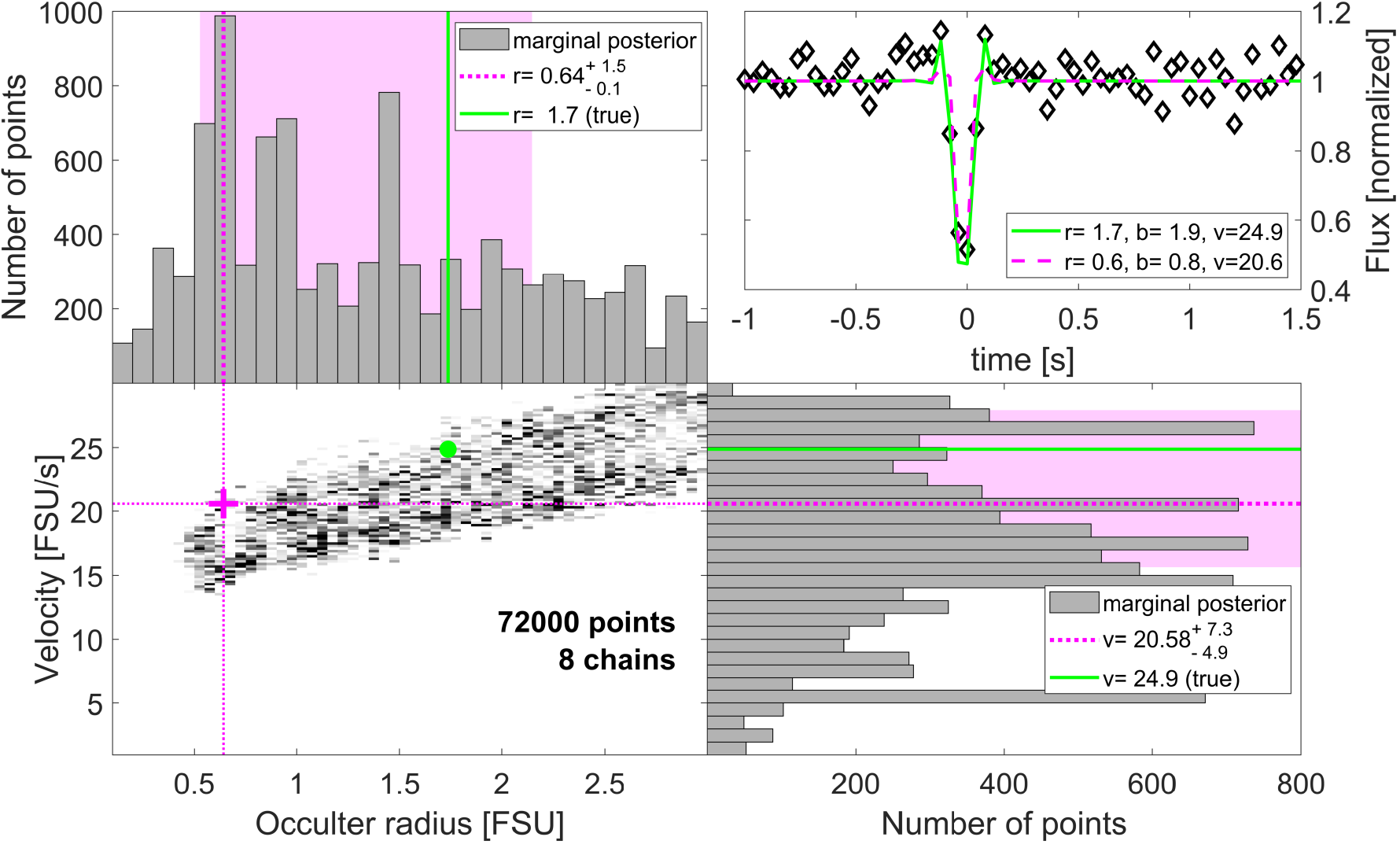}
		\caption{Posterior probability distributions for an event with 
			$R_\star=0.35$\,FSU, $r=1.7$\,FSU, $b=1.9$\,FSU and $v=25$\,FSU\,s$^{-1}$
			(see Figure~\ref{fig: mcmc example chains}). 
			On the lower-left panel we show the density of points
			in the $r$-$v$ space, 
			integrating over all values of $b$ and $R_\star$ (marginalization).
			The 1D histograms (lower-right and upper-left panels) show the marginal posteriors 
			of each individual parameter, $r$ and $v$. 
			The magenta cross and dotted lines represent 
			the best estimate parameters (the mode, see text).
			The best values for this MCMC run are 
			$r= 0.6$\,FSU and $v= 20.6$\,FSU\,s$^{-1}$. 
			The green point and solid lines represent the true point, 
			with the exact parameters used to simulate the injected event. 
			The true values are quite different than 
			the most probable value (the mode)
			but still inside the 68\% bounds. 
			On the top right is the input flux (black diamonds) 
			with two model light-curves for the true point (solid green)
			and the best point (dashed magenta). 
			Even though the parameters for each light-curve are different, 
			the models appear close for most of the data points. 
		}
		\label{fig: mcmc sim results 1}
		
		
	\end{figure*}
	
	\begin{figure*}
		
		\centering
		
		\pic[\mcmcfigsize]{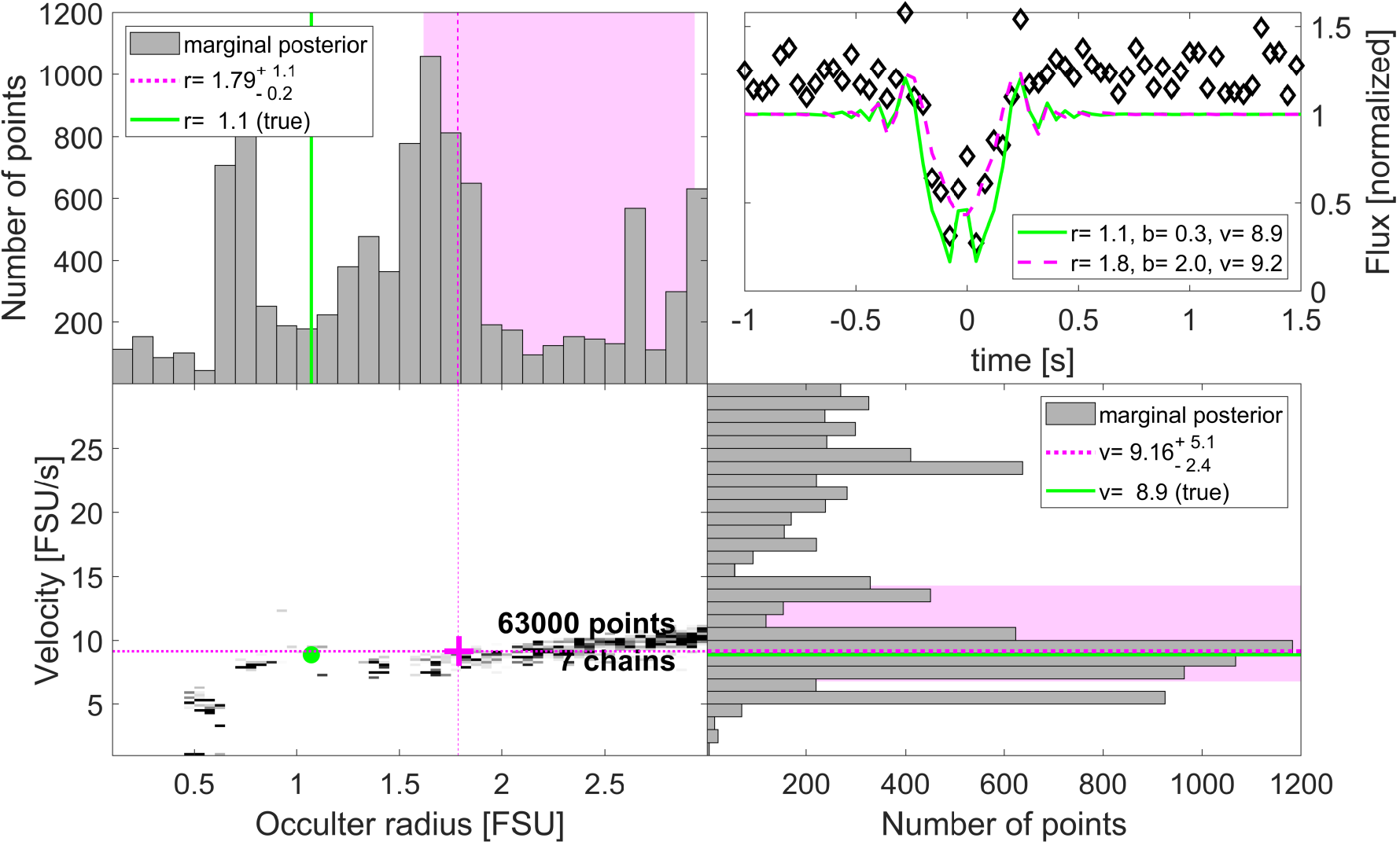}
		\caption{The same as Figure~\ref{fig: mcmc sim results 1}
			only for an event with $R_\star=0.13$\,FSU, $r=1.1$\,FSU, $b=0.3$\,FSU and $v=9$\,FSU\,s$^{-1}$.
			For this event, we find $r=1.8$\,FSU, $b=2.0$\,FSU and $v=9.2$\,FSU\,s$^{-1}$. 
			The lower velocity of the event makes the light-curve very wide, 
			which is hard to confuse with higher velocity values. 
			On the other hand, the occulter radius is still degenerate
			with the impact parameter, 
			allowing the $r$ values to settle on a value
		    much bigger than the true value, 
		    while also choosing a large $b$. 
		}
		\label{fig: mcmc sim results 4}
		
		
	\end{figure*}
	
	\begin{figure*}
		
		\centering
		
		\pic[\mcmcfigsize]{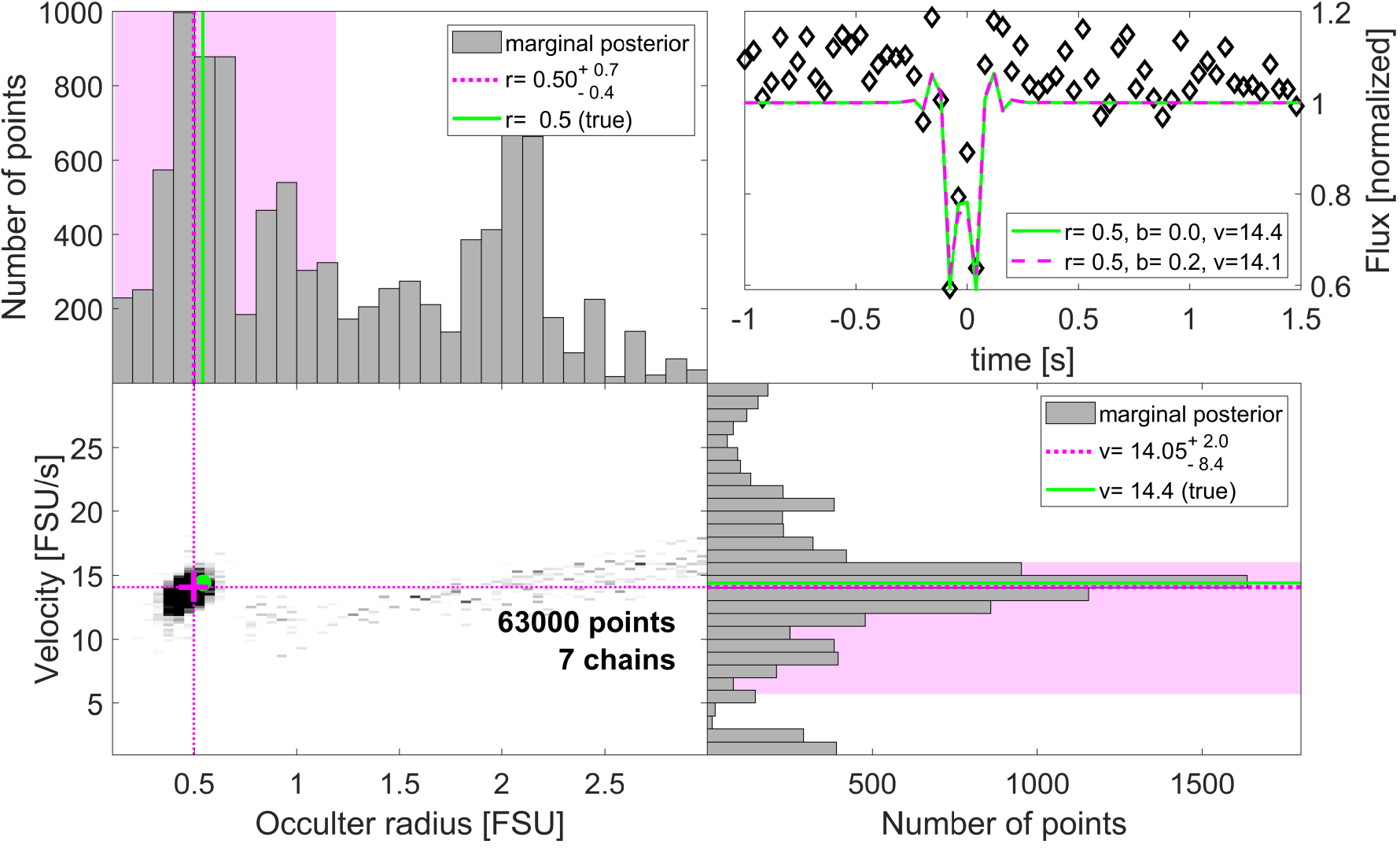}
		\caption{The same as Figure~\ref{fig: mcmc sim results 1}
			only for an event with $R_\star=0.4$\,FSU, $r=0.54$\,FSU, $b=0.01$\,FSU and $v=14.4$\,FSU\,s$^{-1}$.
			For this event, we find $r=0.5$\,FSU, $b=0.2$\,FSU and $v=14.1$\,FSU\,s$^{-1}$. 
			The combination of low velocity and impact parameter, 
			makes distinct shapes in the light-curve.
			The Poisson peak, a brightening in the middle of the occultation, 
			is distinctive enough to break the degeneracy between $r$ and $b$, 
			as it is only seen at low $b$. 
			Most of the points in the MCMC sampler are around the true value, 
			so the distribution is much narrower in all parameters. 
		}
		\label{fig: mcmc sim results 6}
		
		
	\end{figure*}
	
	\begin{figure*}
		
		\centering
		
		\pic[\mcmcfigsize]{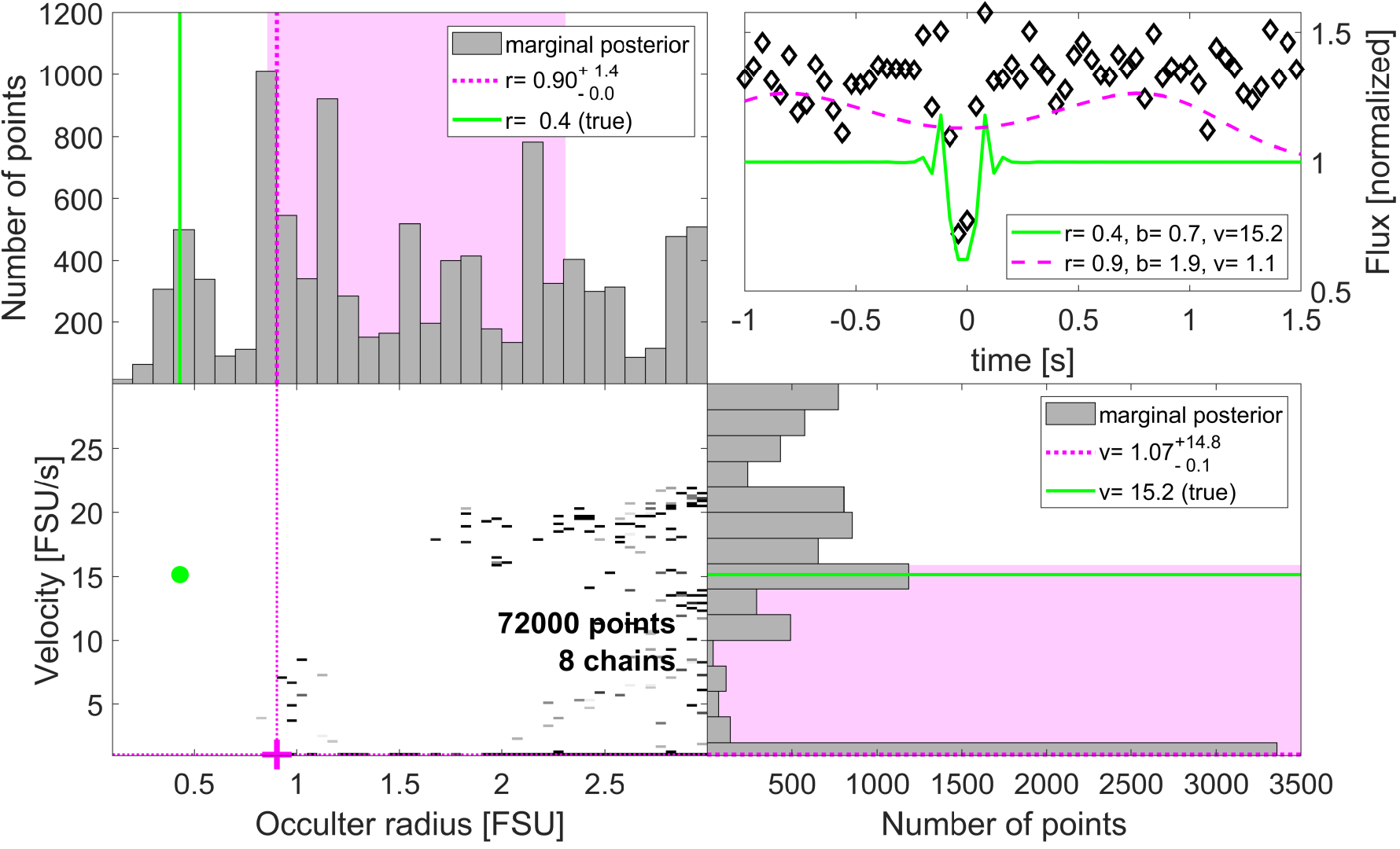}
		\caption{The same as Figure~\ref{fig: mcmc sim results 1}
			only for an event with $R_\star=0.18$\,FSU, $r=0.43$\,FSU, $b=0.7$\,FSU and $v=15$\,FSU\,s$^{-1}$.
			For this event, we find $r=0.9$\,FSU, $b=1.9$\,FSU and $v=1.1$\,FSU\,s$^{-1}$. 
			The sampler seemed to have become stuck in a local minimum, 
			and the results do not reflect the data well. 
			This stresses the need for more random chains
			when analyzing real events. 
			}
		\label{fig: mcmc sim results 3}
		
		
	\end{figure*}
	
	The degeneracy of the occulter radius $r$ and impact parameter $b$, 
	and to some extent the velocity $v$ are evident in most of the sample events. 
	In the first example, shown in Figure~\ref{fig: mcmc sim results 1}, 
	the distribution is fairly wide in both $r$ and $v$, 
	but the range of velocities is more restrictive. 
	
	In cases where the velocity is low ($v\lesssim 10$\,FSU\,s$^{-1}$), 
	the width of the light-curve dip is more restrictive, 
	causing the MCMC chains to be narrow in the $v$ direction. 
	In the $r$ direction, however, there remains a strong degeneracy
	with the impact parameter $b$, 
	such that small $r$ and $b$ are consistent with the data 
	but also large $r$ and $b$, and any intermediate values. 
	This is apparent in the example presented in
	Figure~\ref{fig: mcmc sim results 4}, 
	where the best value of $v$ is very close to the true value, 
	but the value of $r$ is very different. 
	Since in most cases the impact parameter is unconstrained, 
	individual detections will not be able to provide much information
	on the size distribution of KBOs. 
	However, since $b$ is expected to be uniformly distributed 
	(prior to trigger selection bias), 
	it may be possible to infer the distribution of $r$ 
	over a large sample of occultations. 
	Observing the same event with two telescopes
	a few hundred meters apart 
	(a significant fraction of a Fresnel unit, 
	as it is projected to the distance of the occulter)
	may also help break this degeneracy. 
    Observations with a very high S/N (e.g., above 100) could
    begin to resolve the smaller features like the diffraction fringes, 
    which could also help break the degeneracy and narrow down the parameter values. 
    This is, however, difficult to do using ground-based surveys. 
    
	In some cases, where the impact parameter is small ($b\lesssim 0.25$\,FSU), 
	there is a brightening in the middle part of the light-curve, 
	known as the Poisson peak. 
	This peak puts a strong constraint on $b$, 
	breaking the degeneracy and allowing a much tighter handle on $r$. 
	This is only possible when the stellar radius $R_\star$ 
	is not very big ($R_\star \lesssim 0.7$\,FSU). 
	In Figure~\ref{fig: mcmc sim results 6} we show an example 
	for an event with tight constraints on both $v$ and $r$. 
	In this case the velocity ($v\approx 14$\,FSU) 
	and impact parameter ($b\approx 0$\,FSU)
	are low enough to resolve the Poisson peak, 
	and the MCMC indeed samples very close to the true position. 
	
	Finally, in some cases, the particular shape of the 
	light-curve around the occultation, 
	which is determined from random, 
	often correlated noise, 
	can cause the MCMC sampler to 
	get stuck in local minima that 
	are not close to the true shape of the event. 
	In such cases, we see a poor match of the best light-curve, 
	and also scattered MCMC chains that do not seem to 
	converge into any particular area of the parameter space. 
	Such an event is shown in Figure~\ref{fig: mcmc sim results 3}. 
	This highlights the need to run more chains
	at more random starting locations, 
	which minimizes the risk of all chains missing
	the true region of parameters of the event. 
	
	\section{Conclusions}\label{sec: conclusions}
	
	We present the data analysis methodology and pipeline 
	for the detection of KBO occultations in high-cadence photometric data. 
	We discuss methods to deal with red-noise which are common in 
	ground-based observations. 
	We outline a matched-filter-based method and ways to efficiently 
	generate and apply the template bank to large amounts of data. 
	We present a fast simulation tool that can be used for template bank construction, 
	for injection of events into real data, and for parameter estimation. 
	The main differences between this pipeline and previous algorithms
	(e.g., \citealt{TAOS_analysis_Lehner_2010, abundance_kuiper_objects_Schlichting_Ofek_2012, KBO_detection_from_ground_Arimatsu_2019, occultation_real_time_detection_TAOS_II_Huang_2021, Colibri_survey_Mazur_2022})
	are the use of the matched-filtering with a comprehensive template bank, 
	the treatment of correlated noise and the injection of simulated events
	into real data. 
	
	We apply the pipeline to data from W-FAST, 
	a wide-field, fast readout telescope built primarily for detection 
	of occultations by KBOs but also other trans-Neptunian objects. 
	The pipeline reduces imaging data around each star (cutouts)
	to photometric measurements, 
	corrects the light-curves for long-term trends and red-noise, 
	cross correlates (filters) the corrected light-curves
	with occultation templates, 
	taking into account the data power spectral density, 
	and finds candidate occultation events 
	whenever the filtered light-curves exceed the threshold. 
	
	We describe our method for producing simulated occultation templates, 
	which we use for light-curve filtering, 
	for event injection and for parameter estimation of detected events. 
	We use the injected events to estimate the detection efficiency 
	of occultations with various parameter values. 
	
	We describe our method for disqualifying areas of the data with lower quality, 
	estimate the total star-hour yield from typical operations, 
	as well as the number of star-hours lost due to various data quality cuts. 
	We describe the human vetting of candidate events 
	and our blinding scheme for preventing bias in the vetting process. 
	Finally, we describe the MCMC samplers used to estimate the occultation parameters 
	and show the results of such analysis on simulated events. 
	
	This pipeline is being applied to W-FAST data, 
	while future works will discuss the results obtained by processing those datasets, 
	including occultation candidates, 
	and improved limits on the KBO size distribution 
    (e.g., \citealt{WFAST_KBO_search_limits_Nir_2023}). 
		
	\section*{Acknowledgments}
	E.O.O.~is grateful for the support of
        grants from the 
        Willner Family Leadership Institute,
        André Deloro Institute,
        Paul and Tina Gardner,
        The Norman E Alexander Family M Foundation ULTRASAT Data Center Fund,
        Israel Science Foundation,
        Israeli Ministry of Science,
        Minerva,
        BSF, BSF-transformative, NSF-BSF,
        Israel Council for Higher Education (VATAT),
        Sagol Weizmann-MIT,
        Yeda-Sela,
        Weizmann-UK,
        Benozyio center,
        and the Helen-Kimmel center.
        B.Z.~was supported by a research grant from the Willner Family Leadership Institute for the Weizmann Institute of Science

	\section*{Data Availability}

The full data set used in this work, including raw images, calibration files, and the higher level products such as lightcurves and event summaries are available upon reasonable request from the authors. 

	\bibliographystyle{aasjournal}
	
	\bibliography{refs}
	
	\appendix
	
	\section{Description of the quality cuts}\label{sec: quality cuts list}
	
    Besides the flux $f_j(t)$ of each star index $j$ and each measurement time $t$, 
	we also use the auxiliary measurements
	for the centroid position relative to the center of the cutout, $X_j(t)$ and $Y_j(t)$, 
	the PSF width $w_j(t)$ and the background $b_j(t)$ measured in an annulus for each star. 
	Whenever possible, we omit the star and time indices to make notation shorter. 
	In what follows, we use $\overline{\square}$ to denote average over stars, 
	and $\langle \square \rangle$ to denote average over time. 
	
	\begin{itemize}
		\item \code{delta_t} (non-uniform cadence): 
		We search for times when the camera did not maintain a uniform frame rate. 
		We calculate the median difference between frames' timestamps, $\langle \Delta t \rangle$, 
		which represents the average frame rate.  
		This is compared to the measured time difference between adjacent frames, $\Delta t(t)$.  
		We check if there is a large difference between the two, in relative terms:
		\begin{equation}
		  \frac{|\langle \Delta t \rangle - \Delta t(t)|}{\langle \Delta t \rangle} > \tau_{\Delta t}, 
		\end{equation}
		where $\tau_{\Delta t}=0.3$ is the \code{delta_t} threshold. 
		Any points where this value is exceeded are disqualified for all stars. 
		
		\item \code{shakes} (Mean offsets are too large): 
		We calculate the average offset between centroid and center of cutout, 
		calculated over all stars, 
		using the square of the mean flux of each star as weights:
		\begin{align}\label{eq: mean offset}
		\overline{X}(t) &= \sum_j \langle f_j \rangle^2 X_j(t) / \sum_j \langle f_j \rangle^2, \\
		\overline{Y}(t) &= \sum_j \langle f_j \rangle^2 Y_j(t) / \sum_j \langle f_j \rangle^2,
		\end{align}
		Where $\langle f_j \rangle$ are the mean fluxes calculated for each star on that 200 frame period. 
		We use the square of the mean flux to give more weight to bright stars, 
		that have better measurements of first and second moments. 
		
		The two mean offsets, $\overline{X}$ and $\overline{Y}$ are summed in quadrature 
		to give the mean offset size $\overline{R} = \sqrt{\overline{X}^2+\overline{Y}^2}$. 
		When $\overline{R}>\tau_{\overline{R}}=5$ we disqualify all stars in that time range. 
		This is usually caused by strong wind gusts or tracking errors.
		
		\item \code{fwhm} (Mean PSF Full Width at Half Maximum is too high): 
		We take each of the brightest one-hundred stars\footnote{
		Since defocus is consistent between stars in each image, 
		calculating the FWHM only for a subset of the brightest stars 
		helps reduce the computational time, 
		and reaches similar results to using all stars.
		}, 
		and reposition each star in its cutout
		according the centroids $X$ and $Y$ 
		and using FFT partial-pixel shifts (i.e., Shannon interpolation). 
		The shifted cutouts are summed over 200 frames, 
		and the width is calculated 
		by matched-filtering with several templates 
		that describe both focused and defocused stars of various widths. 
		We use (a) a simple 2D Gaussian;
		(b) a generalized Gaussian defined, following \cite{generalized_gaussian_Claxton_Staunton_2008} as
		\begin{equation}
	   	G \propto \exp\left(-\frac{1}{p}\left|\frac{x^2+y^2}{\sigma_{w}^2}\right|^{p/2} \right),
		\end{equation}
		where $p=5$ is the power of the Gaussian; 
		and (c) a ring convolved with a Gaussian with $\sigma_d=1$, 
		representing the shape of a very defocused star. 
		Each of these templates is convolved with the shifted and summed image
		of each star, using different widths $\sigma_{w}$, 
		while normalizing each kernel such that $\sum_i G^2=1$ 
		over all pixels in the kernel. 
		After filtering with $G$, 
		the kernel that has the highest maximal $S/N$ is chosen, 
		and that kernel's $\sigma_w$
		is used as that star's individual width. 
		For the normal and generalized Gaussians, 
		the FWHM is given by $2\sqrt[p]{p \ln 2} \sigma_w$ 
		(which is $2.355 \sigma_w$ for a regular Gaussian
		and $2.56 \sigma_w$ for a generalized Gaussian with $p=5$). 
		In the case of the defocused star ring, the width is given by
		$2r+2.355\sigma_d$, where $r$ is the radius of the ring. 
		We fit the FWHM values of the stars to their $x$ and $y$ position
		in the image using a 2D, second order polynomial. 
		The value of that polynomial at the center of the field 
		is used as the image FWHM estimate. 
	    When this estimate is above $\tau_{\overline{w}}=10''$, 
		the entire batch is disqualified. 
		This sometimes occurs during telescope drifts 
		or at the ends of long observing run when the telescope
		focus drifts over time. 
		
		\item \code{slope} (large slope of the mean flux): 
		We search for frames where the mean flux of all stars changes rapidly over time
		(either ramps up or slopes down). 
		We use a slope kernel with time-scale $t_s$:
		\begin{equation}\label{eq: slope kernel}
    		K_s(t) = \begin{cases} t & |t|<t_s/2 \\ 0 & \text{otherwise} \end{cases}. 
		\end{equation}
		This filter is normalized so that $\sum_t K_s^2(t)=1$. 
		For this cut we use a time-scale of $t_{s, \text{slope}}=50$ frames. 
		We calculate the average flux $\overline{f}$ over the all stars, 
		and normalize it by its mean and standard deviation within the 200 frame region:
		\begin{equation}
		  \overline{f_N} = \frac{\overline{f} - \langle \overline{f} \rangle}{\text{STD} (\overline{f})}
		\end{equation}
		The kernel $K_s$ is used to filter the normalized flux $\overline{f_N}$, 
		and we disqualify all stars whenever 
		the filtered flux surpasses the threshold $\tau_\text{slope}=5$:					
		\begin{equation}
		  K_s(t) \star \overline{f_N} > \tau_\text{slope},
		\end{equation}
		This cut is triggered when, e.g., thin clouds affect the global image transparency. 
		
		\item \code{instability} (Mean flux over time changes too rapidly): 
		For each star we calculate the Median Absolute Deviation (MAD) 
		for the full flux buffer (typically 2000 samples), 
		and compare that with the standard deviation of each bin of 100 samples. 
		If the ratio is larger than $\tau_\text{instability} = 3$, 
		that region of the data is disqualified. 
		This cut is sensitive to visual binaries seen in the same cutout, 
		or to slow but dramatic changes in visibility or background levels. 
		
		\item \code{near_bad_rows_cols} (star centroid is too close to a bad row/column):
		If at any time the star centroid is closer than five pixels 
		to any of the known bad rows or columns on the sensor, 
		that star is disqualified during that time. 
		
		\item \code{bad_pixels} (bad pixels included in the area of the photometric aperture): 
		We disqualify any frames in which there are any bad pixels inside the photometric aperture. 
		
		\item \code{offset_size} (star centroid offsets are too big):
		After calculating the mean offsets $\overline{X}$ and $\overline{Y}$ in each frame, 
		we can calculate the residual offsets $x=X-\overline{X}$ and $y=Y-\overline{Y}$. 
		These represent the individual star's motion relative to the global motion 
		of the telescope against the sky. 
		These motions are due to a mixture of atmospheric scintillation, 
		satellites passing through the cutouts, and other artefacts. 
		The same residual offsets are used in this and the following quality cuts. 
		If the size of the offsets for a given star, 
		$r=\sqrt{x^2+y^2}$, is larger than $\tau_r=4$ for three or more frames in a row, 
		we disqualify all frames except the first and last. 
		
		Since the centroids often have large fluctuations, 
		especially for faint stars, 
		we require at least three consecutive frames above threshold to trigger this cut. 
		This dramatically reduces the number of star hours lost to this cut, 
		while maintaining its efficiency against satellites and image artefacts. 
		
		\item \code{linear_motion} (the offsets change in a linear way): 
		We search for stars that show a linear change in the residual centroid offsets $x$ and $y$. 
		To look for such occurrences we filter the $x$ and $y$ auxiliary measurements 
		with a slope kernel $K_s$ as defined in Equation~\ref{eq: slope kernel}, 
		with a time-scale of $t_{s, \text{motion}} = 25$ frames. 
		
		For each star we filter the (mean subtracted) offset in $x$ and $y$ 
		with the slope kernel:
		\begin{align}
		  S_x(t) &= K_s \star (x(t) - \langle x \rangle), \\
		  S_y(t) &= K_s \star (y(t) - \langle y \rangle), 
		\end{align}
		where $\langle x \rangle$ and $\langle y \rangle$ are the 
		time-averaged offsets over the 200 frames, for each star individually. 
		These values are used primarily to disqualify satellites crossing the image cutouts. 
		Thus, we only apply this cut to frames where the flux is unusually high. 
		We calculate the normalized flux, 
		\begin{equation}
		  f_\text{norm} = \frac{f-\langle f \rangle}{\text{STD}(f)},
		\end{equation}
		where STD($f$) is the standard deviation of the flux in those 200 frames. 
		We calculate the moving average of the normalized flux over the timescale $t_{s, \text{motion}}$:
		\begin{equation}
		  f_\text{average} = \sum_{t=-t_s/2}^{t_s/2} f_\text{norm}(t) / N_{t_s},
		\end{equation}
		where $N_{t_s}$ is the number of frames in the time-scale $t_s$. 
		For any star we disqualify any time that has 
		\begin{equation}
		  L(t) = \max_{t_s} \left[f_\text{average} \sqrt{S_x^2 + S_y^2}\right] >\tau_L,
		\end{equation}
		where $\tau_L=2$ is the threshold for the linear motion cut, 
		and $\max_{t_s}$ is the maximum value of the cut in a moving time window 
		of width $t_s$. 
		The moving maximum helps disqualify events where the 
		linear motion values are high right before or after the closest approach 
		of the satellite to the star, 
		but not directly overlapping with the event trigger. 
		
		\item \code{background_intensity} (background levels too high):
		we compare the value of the background measurements taken from 
		averaging the pixel values in the photometric annulus 
		(see Appendix~\ref{sec: photometry})
		to a fairly high threshold of $\tau_\text{b/g} = 10$. 
		For the short exposures used in this analysis, 
		it is very rare that the background in individual images
		surpasses one or two counts per pixel. 
		This cut usually disqualifies passing satellites 
		that go through a star's annulus, or during twilight. 
		
		\item \code{aperture_difference} (fluxes from different photometric apertures are different): 
		We compare the flux measured using two photometric apertures with the same radius, 
		but with one centered around the star's centroids (denoted $f_\text{unforced}$), 
		and the other centered around the (flux-weighted) average centroids 
		measured using all the stars
		which we refer to as \emph{forced aperture}, denoted as $f_\text{forced}$,
		which are the fluxes we use throughout this pipeline
		(see Appendix~\ref{sec: photometry}). 
		We calculate the relative difference between the two fluxes:
		\begin{equation}
		    \frac{f_\text{forced} - f_\text{unforced}}{\text{MAD}(f_\text{forced})} > \tau_\text{ap.diff.}
		\end{equation}
		and disqualify values that surpass the threshold $\tau_\text{ap.diff.}=4$. 
		This cut helps remove false detections where a drift between the centroid 
		of a single star relative to the average centroid, 
		causes the forced photometry to show dips in the light-curve. 
		Since true occultations should appear in either aperture position, 
		we disqualify times when the two do not match. 
		
		\item \code{corr_x_25}, \code{corr_w_50}, etc.~(correlation of flux with auxiliary measurements): 
		We use the cross-correlation coefficient defined over two time-series $a$ and $b$: 
		\begin{equation}\label{eq: correlation coefficient}
		  C(a,b) = \frac{\sum (a-\langle a \rangle)(b - \langle b \rangle)}  {\sqrt{ \sum (a - \langle a \rangle)^2 \sum (b - \langle b \rangle)^2}}, 
		\end{equation}
		where the sums are over a certain duration of the time-series. 
		For our calculations, we use a rolling window over the data, 
		with windows of time-scale 25 and 50 frames. 
		As the time-scale grows larger, the average correlation values
		for uncorrelated time-series decreases like the square root of the time-scale. 
		To make all these correlation values follow the same baseline 
		we normalize by multiplying each correlation value 
		by the square root of the number of frames in each window. 
		
		The correlations we test are those of the raw flux $f$, 
		with a few of the auxiliary values. 
		Namely we correlate the flux with the backgrounds $b$, 
		the residual offsets $x$ and $y$, 
		the size of these offsets $r$, 
		and the PSF width $w$. 
		These five auxiliaries, with the two time-scales, 
		give us 10 different cuts on the data.  
		We disqualify any frames and stars where any of these correlations
		surpass the threshold $\tau_\text{corr}=4$. 
		
		\item \code{flux_corr_25} and \code{flux_corr_50} (correlations between stars' fluxes):
		We calculate the correlation (see Equation \ref{eq: correlation coefficient}) 
		between every star's flux with each of the brightest 1000 stars.\footnote{ 
		As this cut is targeted at times when the flux is affected by global influences, 
		correlating against the brightest stars is good enough to 
		estimate if there are any significant flux correlations. 
		}
		This is calculated at a few times in each batch, 
		with intervals of half the correlation length 
		(i.e., at 12 and 25 frames for correlations of 25 and 50 frames, respectively). 
		For each sampling point, each star's flux is correlated with 
		the brightest 1000 stars and the results 
		are multiplied by the squre root of the correlation length. 
		Then we sort those values and choose the 20th most correlated flux. 
		This avoids flux correlations between specific stars, 
		and gives a better estimate of the strength of the global flux correlations. 
		The flux correlation values are interpolated between the sampling points
		using a linear interpolation to find the estimated correlation
		values for any frame in the batch. 
		We disqualify any frames and stars where any of the flux correlations
		surpass the threshold $\tau_\text{flux corr}=4$. 
		
	    \item \code{repeating_columns} (data is corrupt with duplicated pixel values): 
	    In a small fraction of the data, camera software issues 
	    cause values in columns (or parts of columns) in one image
	    to be copied into images taken exactly 10 frames later. 
	    We disqualify any cutout where pixels along the top or bottom half of any column 
	    in any cutout have the same values as the corresponding pixels 
	    in an image taken 10 frames before that.

	\end{itemize}
	
	Each of the above cuts is applied to each star 
	in each frame of the 200 frame section of the data. 
	We use 200 frames for the various filtering operations, 
	but only use the data in the central 100 frames to disqualify times, 
	as that is the region we use for event finding. 
	
	Some of these cuts apply for all stars at the same time
	(the \code{delta_t}, \code{shakes}, and \code{slope} cuts). 
	The rest can affect each frame and each star individually. 
	It is common for these cuts to affect continuous regions, 
	particularly when using large time-scale filtering kernels, 
	and in some cases entire batches of data are disqualified for certain stars
	(as is common for the \code{near_bad_rows_cols} cut).

	\section{Photometry} \label{sec: photometry}
	
	Cutouts are dark and flat corrected, and fed into a fast custom photometry code. 
	This code uses simple aperture photometry, 
	as well as tapered apertures 
	where the weight of each pixel decreases with distance from the center
	of the tapered aperture by a Gaussian function:
	\begin{equation}
	   G_i = \exp\left(\frac{(x_i-x_c)^2+(y_i-y_c)^2}{2\sigma_g^2}\right), 
	\end{equation}
	Where $x_i$ and $y_i$ are the pixel coordinates, 
	$x_c$ and $y_c$ are the center of the mask position,  
	and $\sigma_g$ is the Gaussian width parameter, 
	which we set to 2 pixels in our analysis.
	The Gaussian aperture is used to find the first and second moments 
	of the star's Point Spread Function (PSF), 
	as it is less sensitive to noise. 
	
	Each cutout is first subjected to simple photometry, 
	where the flux and centroid is measured using all the pixels
	without any masking. 
	The background subtraction is done using an annulus centered
	at the center of the cutout.
	Then an aperture is centered around the centroid found in the unmasked image. 
	An annulus is centered around the same point. 
	The largest aperture has a radius of 7 pixels ($16''$), 
    but flux is also calculated using concentric apertures of 
    3 pixels ($6.9''$) and 5 pixels ($11.45''$). 
    In most cases we use the flux using the smallest aperture 
    for occultation searches, as they tend to have the lowest noise. 
    In some cases, one of the larger apertures could provide more stable
    photometry (e.g., if the PSF is very wide or if its width changes drastically between images). 
    In practice, this has not been used in any of the W-FAST data so far. 
    
    The annulus used for all measurements has an inner and outer radius
	of 7.5 and 10 pixels ($17.25''$ and $23''$). 
	The annulus is used to subtract the background 
	and once again a new centroid is located. 
	This centroid is used as the center point for a Gaussian mask 
	and a new annulus, that are used to produce even better centroid positions. 
	Finally, a second iteration of a Gaussian mask using the latest centroid 
	is used to find the centroids and width of the PSF based on the second moments.
	The first moments are calculated using
	\begin{equation}
	   M_{x} = \sum_i x_i (I_i-B)G, \quad M_{y} = \sum_i y_i (I_i-B)G,
	\end{equation}
	while the second moments are calculated using:
	\begin{align}
	   M_{xx} &= \sum_i x_i^2 (I_i-B)G, \quad M_{yy} = \sum_i y_i^2 (I_i-B)G, \notag \\ 
	   M_{xy} &= \sum_i x_i y_i (I_i-B)G,
	\end{align}
	where $x_i$ and $y_i$ are the pixel coordinates in the cutout, 
	$I_i$ are the pixel values, $B$ is the background calculated 
	using the repositioned annulus, 
	and $G$ is a Gaussian mask that is iteratively repositioned using the previous iteration's best centroids. 
	The second moments are translated into the major and minor axis of an ellipse 
	using the matrix inversion method:
	\begin{align}
	   \text{tr} &= M_{xx}+M_{yy}, \quad \text{det}=M_{xx}*M_{yy} - M_{xy}^2 \notag \\
	   r_{\pm}   &= \sqrt{\text{tr} \pm \sqrt{\text{tr}^2 -4\text{ det}}}
	\end{align}
	where $r_\pm$ are the major and minor axes. 	
	The average of the two axes is taken as the width of the PSF, 
	$\sigma_\text{psf,gauss}=(r_{-} + r_{+})/2$, 	 
	which is translated to Full Width Half Maximum (FWHM) by 
	multiplying by a factor of 2.355. 
	Since we are using the measurement of the width from a tapered aperture, 
	we must take into consideration the narrowing caused by the weighting processes. 
	The real width is recovered using:
	\begin{equation}
	   w = \frac{\sigma_\text{psf,gauss}\sigma_g}{\sqrt{\sigma_g^2 - \sigma_\text{psf,gauss}^2}}.
	\end{equation}

	Along with the width $w$, we also calculate $X=M_x$ and $Y=M_y$, 
	the centroid positions of each star in each cutout (also based on the Gaussian-tapering aperture), 
	the background $b$ based on the annulus 
	and flux $f$ based on the circular (un-tapered) aperture. 
	We refer to all measurements besides the flux 
	as `auxiliary' photometric products.

\end{document}